\documentclass[12pt]{iopart}
\usepackage{iopams}
\expandafter\let\csname equation*\endcsname\relax
\expandafter\let\csname endequation*\endcsname\relax
\usepackage{amsmath}
\usepackage{graphicx}
\usepackage{caption}
\usepackage[usenames,dvipsnames,svgnames,table]{xcolor}
\begin{document}

\title[Evolutionary accessibility of random and structured fitness landscapes]{Evolutionary accessibility of random and structured fitness landscapes}

\author{Joachim Krug and Daniel Oros}

\address{Institute for Biological Physics, University of Cologne, Z\"ulpicher Strasse 77, 50937 K\"oln, Germany}
\ead{jkrug@uni-koeln.de}

\begin{abstract}
Biological evolution can be conceptualized as a search process in the space of gene sequences guided by the fitness landscape, a mapping that assigns a measure of reproductive value to each genotype. Here we discuss probabilistic models of fitness landscapes with a focus on their evolutionary accessibility, where a path in a fitness landscape is said to be accessible if the fitness values encountered along the path increase monotonically. For uncorrelated (random) landscapes with independent and identically distributed fitness values, the probability of existence of accessible paths between genotypes at a distance linear in the sequence length $L$ becomes nonzero at a nontrivial threshold value of the fitness difference between the initial and final genotype, which can be explicitly computed for large classes of genotype graphs. 
The behaviour in uncorrelated random landscapes is contrasted with landscape models that display additional, biologically motivated structural features. In particular, landscapes defined by a tradeoff between adaptation to environmental extremes have been found to display a combinatorially large number of accessible paths to all local fitness maxima. We show that this property is characteristic of a broad class of models
that satisfy a certain global constraint, and provide further examples from this class.
\end{abstract}

%
%
%
%
%

\vspace*{0.5cm}

\hspace*{0.5cm} \hfill \textit{Klettere, steige, steige. Aber es gibt keine Spitze.}\footnote{Climb, ascend, ascend. But there is no summit.} \\
\hspace*{0.5cm} \hfill Kurt Tucholsky
\normalsize

\section{Introduction}

\begin{figure}[htb]
    \centering
    \includegraphics[width=0.75\textwidth]{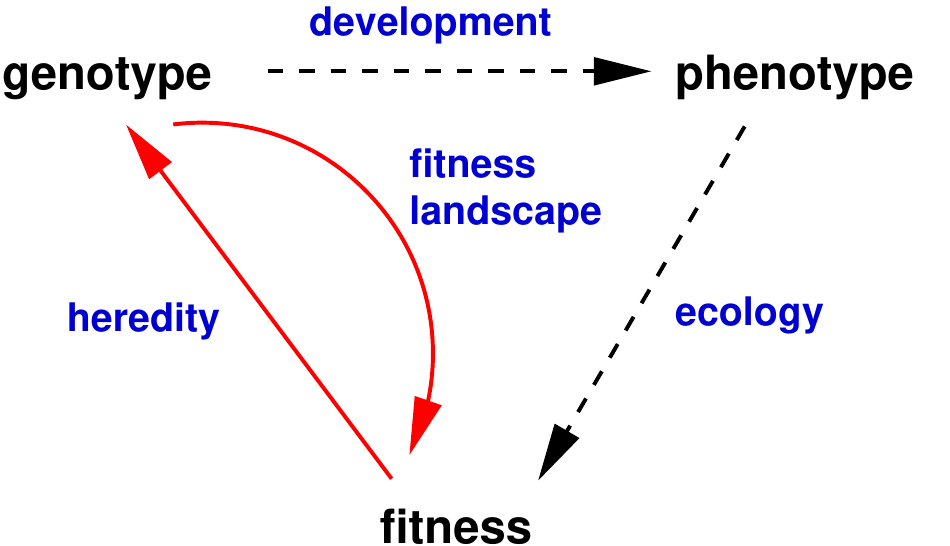}
    \caption{\textbf{The fitness landscape concept in evolutionary theory.} Darwinian evolution is driven by a feedback cycle
      consisting of three major steps. Through the process of \textit{development}, the physical features of the organism
      (its \textit{phenotype}) emerge from the
      genotype. \textit{Ecological interactions} of the phenotype with the abiotic and biotic environment determine the organisms reproductive success (its \textit{fitness}). Finally, by the phenomenon of \textit{heredity}, the reproductive success determines the relative abundance of genotypes in the next generation. The fitness landscape constitutes a conceptual shortcut where
      the complexities of development and ecology are bypassed. The
      figure was inspired by a lecture by Amitabh Joshi at the Jawaharlal Nehru Centre for Advanced
      Scientific Research in Bangalore in 2015.
    }
    \label{fig:fig1}
\end{figure}

The description of biological evolution in terms of a fitness landscape was introduced by Sewall Wright almost
a century ago \cite{Wright1932}. The  underlying conceptual step is as simple as it is bold: Given that evolution is
driven by heritable variation within a population, where the genes of
individuals with better reproductive capability (\textit{fitness}) spread by natural
selection, it is tempting to consider a theoretical framework that maps
the genetic makeup of an organism (its \textit{genotype}) directly to
its fitness. Needless to say, this is an enormous
oversimplification, by which much of the structure of the
biological world is swept under the rug (figure \ref{fig:fig1}). Nevertheless the
fitness landscape concept has proved to be of great heuristic value,
and it continues to play a central role in current evolutionary theory
\cite{Svensson2012,deVisser2014,Hartl2014,Kondrashov2015,Fragata2019,Manrubia2021,Bank2022}.

Already in his seminal article Wright expressed the concern that the
\textit{ruggedness}  of the fitness landscape, that is, the existence of
multiple fitness peaks, might compromise the ability of natural selection to
access optimal evolutionary solutions: 

\begin{itemize}

\item[]   
\textit{In a rugged field of this character, selection will easily carry the species to the
nearest peak, but there will be innumerable other peaks that will be higher but
which are separated by ``valleys''. The problem of evolution as I see it is that
of a mechanism by which the species may continually find its way from lower
to higher peaks in such a field.} \cite{Wright1932}

\end{itemize}
Wright's expectation that real fitness landscapes possess
\textit{``innumerable other peaks''} was not shared by everyone,
however. In a letter written in 1931, Ronald Fisher speculated that
true fitness peaks should in fact be very rare, because
most stationary points of a function on a high-dimensional space are
saddles rather than maxima or minima \cite{Provine1986}. Similar
arguments can be found in the recent literature \cite{Agarwala2019}, and the question
to what extent our view of the evolutionary process is misled 
by intuition gained from low-dimensional landscapes is still under
debate \cite{Greenbury2022}.

In this contribution we address the evolutionary accessibility of
high-dimensional rugged fitness landscapes within a somewhat
restricted, but mathematically clearly defined setting. Formalising
the fitness landscape as a real-valued function on the space of
genetic sequences, we ask for the probability that two of these
sequences are connected by a mutational path along which the fitness
function increases monotonically
\cite{Weinreich2005,Poelwijk2007,Carneiro2010,Franke2011}. This notion of
accessibility was first formulated by Daniel Weinreich and
collaborators in their pioneering studies of small combinatorially
complete fitness landscapes describing the evolution of antimicrobial
drug resistance \cite{Weinreich2006,DePristo2007,Lozovsky2009}.

Meanwhile the scale
of the empirical fitness landscapes that can be explored using
state-of-the-art high throughput methods has increased enormously
\cite{Palmer2015,Bank2016,Wu2016,Aguilar2017,Domingo2018,Pokusaeva2019,Moulana2022,Papkou2023,Westmann2023}, but
the factors determining how accessible these landscapes are remain
largely unknown. For example, in an analysis of the affinity landscape
spanned by the 15 point mutations separating the spike protein of the ancestral Wuhan
strain of the SARS-CoV2 virus from the Omicron BA.1 variant, it was found that none of the $15! \approx 1.3 \times 10^{12}$ direct
mutational pathways\footnote{See section \ref{Sec:Space} for the definition of
  direct paths.} is monotonically increasing \cite{Moulana2022}. On
the other hand, a study of combinations of nucleotides at 9
positions of the gene coding for the dihydrofolate reductase (DHFR) enzyme in \textit{Escherichia
  coli} subjected to the antibiotic trimethoprim revealed a fitness
landscape that combines high ruggedness with high accessibility \cite{Papkou2023}; we
will return to this example below in Sect.~\ref{Sec:highly}.

The purpose of this article is to provide a concise and coherent account of our current
understanding of evolutionary accessibility in probabilistic models of fitness landscapes. We distinguish between \textit{random}
(uncorrelated) landscapes, where fitness values assigned to genotypes are independent and identically distributed (i.i.d.)
random variables, and \textit{structured} landscapes that possess some kind of (more or less biologically motivated)
fitness correlation. The next section introduces the mathematical setting, and explains in particular the description of
sequence space as the Cartesian power of a mutation graph \cite{Schmiegelt2023}. Known results for random and structured
landscapes are reviewed in sections \ref{Sec:random} and \ref{Sec:structured}, respectively. An important lesson from
the work on structured landscapes is that ruggedness does not generally correlate with accessibility in a simple way. 
Section \ref{Sec:highly} describes a recently discovered class of fitness landscape models that combine high accessibility
with high ruggedness. Most of the material in this section has not been reported previously. The article concludes with
a summary and an outlook in section \ref{Sec:Conclusions}.

\section{Sequence space and accessible paths}
\label{Sec:Space}

A genotype is a sequence $\sigma = (\sigma_1, \dots, \sigma_L)$ of
length $L$ with entries $\sigma_i$ drawn from an alphabet of size $a$,
which we represent by the numbers $0, \dots, a-1$. In the following
the entries in the sequence will be referred to as \textit{alleles}, and the positions $i = 1,\dots,L$ in the sequence
as \textit{loci}. The set
$\{0,\dots,a-1\}^L$ equipped with the natural distance measure
\begin{equation}
  \label{eq:Hamming}
  d(\sigma,\tau) = \sum_{i=1}^L 1 - \delta_{\sigma_i, \tau_i}
\end{equation}
defines the \textit{Hamming graph} $\mathbb{H}_{a}^L$ \cite{Stadler1999}, where genotypes at mutational distance $d=1$
are connected by a link. The
binary Hamming graph $\mathbb{H}_2^L$ is the $L$-dimensional discrete
hypercube.

\begin{figure}[htb]
  \centering
    \includegraphics[width=\textwidth]{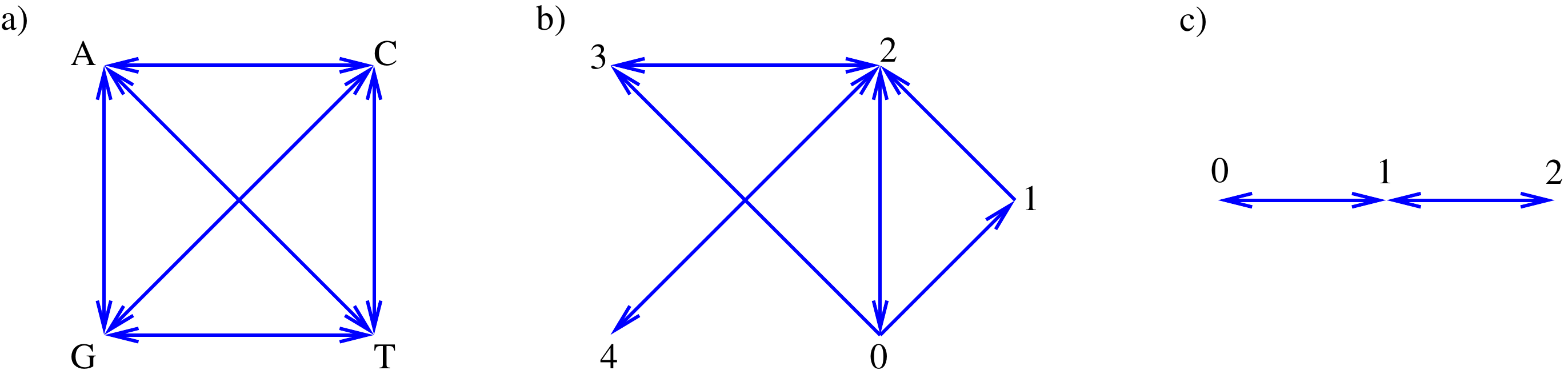}
    \caption{\textbf{Examples of allele graphs.} a) The allele graph
      of the nucleotide alphabet of DNA-sequences is the complete
      graph over 4 alleles. b) An incomplete graph over 5 alleles. This graph is the smallest known example that admits an
      accessibility problem of irregular type. In this case the paths
      are assumed to move from 
      $\alpha = (0 \dots 0)$ to $\omega = (4 \dots 4)$.
      See section \ref{sec:undirected} for details. c)
      The linear path graph over 3 alleles. Such a graph appears
      naturally if the different alleles represent the copy number of a
      gene, which changes by deletion or duplication \cite{Altenberg2015}. 
    }
    \label{fig:allelegraphs}
\end{figure}

The description of sequence spaces by Hamming graphs assumes that any allele can mutate to any other allele, which is not
always the case. For example, the genetic code restricts the possible single nucleotide mutations that convert one amino acid
into another. To account for this, it is useful to introduce the \textit{allele graph} ${\mathcal{A}}$ over the allele
set $\{0, \dots, a-1\}$, which encodes the allowed mutational transitions \cite{Schmiegelt2023}. The adjacency matrix
$\mathbf{A} = \{A_{\mu \nu}\}_{\mu,\nu=0,\dots,a-1}$ of ${\mathcal{A}}$ has an entry $A_{\mu \nu} = 1$ iff the mutation $\mu \to \nu$
is possible,
and $A_{\mu \nu} = 0$ otherwise. Examples of allele graphs are shown in figure \ref{fig:allelegraphs}. The sequence space is then
the $L$-fold Cartesian product of ${\mathcal{A}}$, which we denote by ${\mathcal{A}}^L$. The Hamming graph $\mathbb{H}_{a}^L$
is the $L$-fold product of the complete graph over $a$ alleles.

\begin{figure}[htb]
    \centering
    \includegraphics[width=0.4\textwidth]{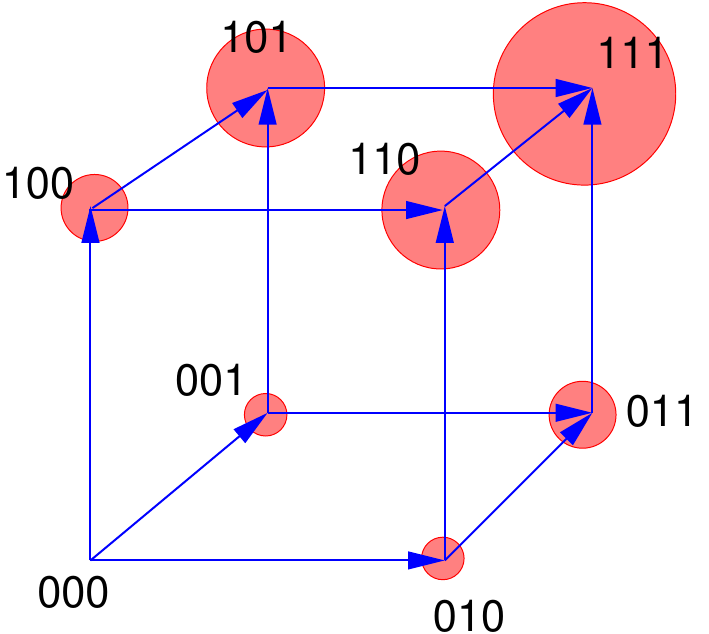}\hspace*{1.5cm}\includegraphics[width=0.4\textwidth]{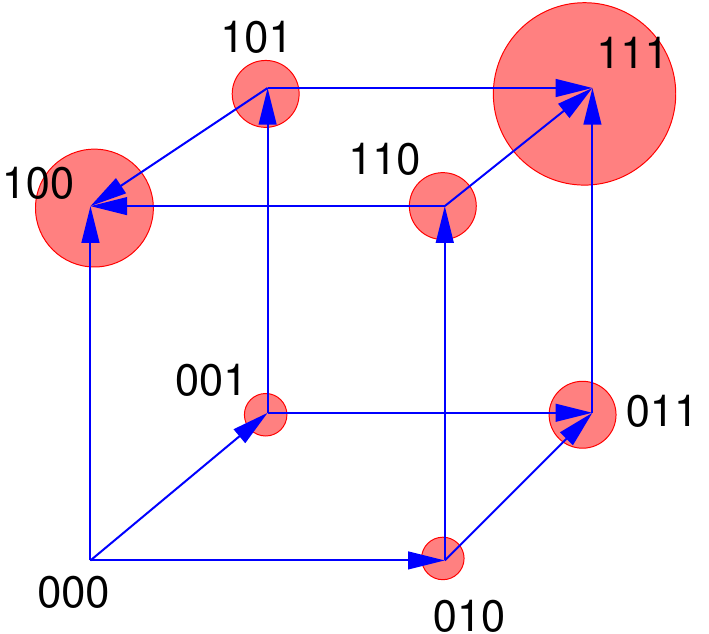}
    \caption{\textbf{Fitness graphs on the 3-cube.} The figure shows
      two examples of fitness landscapes on the binary 3-cube. Fitness
      values are represented by the size of the circles on the nodes,
      and the arrows on the links point in the direction of increasing
      fitness. The landscape in the left panel has a unique maximum at
      $\sigma = (1,1,1)$, and there are 6 direct accessible paths of minimal
      length 3 from $(0,0,0)$ to $(1,1,1)$. In the right panel there
      is a local maximum at $(1,0,0)$ and a global maximum at
      $(1,1,1)$. As a consequence, there are now 4 direct accessible paths $(0,0,0) \to
      (1,1,1)$, and one direct and two indirect accessible paths
      $(0,0,0) \to (1,0,0)$. Each of the two indirect accessible
      paths contains a backstep (mutational reversion) $1 \to 0$. 
    }
    \label{fig:fig2}
\end{figure}

A fitness landscape is a real-valued function
\begin{equation}
  \label{eq:landscape}
  g:  {\mathcal{A}}^L \to \mathbb{R}.
\end{equation}
By decorating the links by arrows pointing in the
direction of increasing fitness, the fitness landscape induces an
\textit{orientation} on the graph which by construction is
acyclic (figure \ref{fig:fig2}) . The resulting oriented graph is called a \textit{fitness
  graph} \cite{deVisser2009,Crona2013}.

In order to make the fitness graph construction unambiguous we will assume throughout that $g$ is non-degenerate, such that no two
genotypes have exactly the same fitness. While this condition is naturally
satisfied in probabilistic models where fitness values are drawn from a
continuous distribution, it may seem to ignore the important role of
neutral mutations \cite{Manrubia2021,Greenbury2022}. In this context it should be
noted, however, that, empirically, not
even synonymous mutations (i.e., mutations that change the nucleotide in the DNA
sequence but not the corresponding amino acid) are strictly neutral
\cite{Plotkin2011,Zwart2018}.

A \textit{mutational pathway} on ${\mathcal{A}}^L$ is a sequence of genotypes
\begin{equation}
  \label{eq:path}
  \alpha = \sigma^{(0)}  \to \sigma^{(1)} \to \dots \to \sigma^{(\ell)} = \omega
\end{equation}
where $d(\sigma^{(i)}, \sigma^{(i+1)}) = 1$ for $i = 0,\dots,\ell-1$, and all mutational steps are allowed by the allele graph.
A path is called
\textit{accessible} if fitness increases in each step, 
\begin{equation}
  g(\sigma^{(i+1)}) > g(\sigma^{(i)})
\end{equation}
for all $i$. In population genetic terminology, all mutational steps are \textit{beneficial}, whereas \textit{deleterious}
mutations (which decrease fitness) do not occur. In other words, accessible paths respect the orientation of the
fitness graph. Because the orientation is acyclic, accessible paths
are self-avoiding. A path is called \textit{direct} if
\begin{equation}
  \label{eq:directed}
  d(\alpha,\sigma^{(i)}) = i,
\end{equation}
which implies that each step reduces the distance to the target genotype $\omega$ by one. Paths which violate
(\ref{eq:directed}) are \textit{indirect}. 
 
\section{Random fitness landscapes:
  Accessibility percolation}
\label{Sec:random}

The most basic probabilistic fitness landscape is the
House-of-Cards (HoC) model, where fitness values assigned to genotypes
are independent, identically and continuously distributed random
variables \cite{Kingman1978,Kauffman1987}. The HoC model is the
conceptual analogue of the Random Energy Model of spin glasses
\cite{Derrida1981}. Since the fitness graph is fully determined by the
rank ordering of the fitness values \cite{Crona2017}, the statistics
of accessible paths are independent of the choice of the fitness
distribution, provided it is continuous. For notational convenience,
we will nevertheless assume that the fitness values are uniformly
distributed on the unit interval $[0,1]$.

For a pair of genotypes $\alpha, \omega$ on the HoC
landscape, the number of accessible paths from $\alpha$ to $\omega$ is a
non-negative integer valued random variable $X_{\alpha,\omega}$, and we
say that $\omega$ is accessible from $\alpha$ if $X_{\alpha,\omega} \geq
1$ \cite{Franke2011}. The probability for this event obviously depends on the fitness
difference
\begin{equation}
  \label{eq:beta}
  \beta = g(\omega) - g(\alpha)
\end{equation}
  and on the distance $d(\alpha,\omega)$. In this section we are
concerned with the behaviour of the probability
$\mathbb{P}[X_{\alpha,\omega} \geq 1]$ as a function of $\beta$ for paths with
$d(\alpha,\omega) \sim L$ when $L \to \infty$. In many cases (to be
specified below) this probability displays a sharp transition from 0
to a nonzero value at a threshold $\beta_c$, a phenomenon that has
been referred to as \textit{accessibility percolation}
\cite{Nowak2013}. Since comprehensive recent accounts of this topic
are available \cite{Schmiegelt2023,Krug2021,Schmiegelt2023a}, we restrict
ourselves to a summary of the main ideas and results.

\subsection{Direct paths}
\label{sec:directed}

Consider a path (direct or indirect) of length $\ell$ that connects two genotypes $\alpha$
and $\omega$ with a fitness difference $\beta > 0$. The path is accessible
if (i) the $\ell - 1$ fitness values of the intermediate genotypes
$\sigma^{(1)}, \dots \sigma^{(\ell-1)}$ are in the interval
$(g(\alpha),  g(\omega))$, which is true with
probability $\beta^{\ell-1}$ and (ii) these fitness values are
increasingly ordered, which is true with probability
$\frac{1}{(\ell-1)!}$. The joint probability of the two events is 
\begin{equation}
\label{eq:Pbetaell}
P_{\beta,\ell} = \frac{\beta^{\ell-1}}{(\ell -1)!}.
\end{equation}
To highlight the relation to percolation phenomena, we provide an alternative interpretation of this
expression \cite{Schmiegelt2023,Krug2021}. Suppose that we fix the fitness values of the path
endpoints at their extremal values, $g(\alpha) = 0$ and
$g(\omega) = 1$, and remove all other genotypes independently
with probability $1-\beta$. Then the probability for existence of an
accessible path \textit{on extant genotypes} is again given by (\ref{eq:Pbetaell}), where the
factor $\beta^{\ell-1}$ is now the probability that none of the
intermediate genotypes have been removed.

For a direct path the sequences $\alpha = (\alpha_1, \dots, \alpha_L)$ and $\omega = (\omega_1, \dots, \omega_L)$
differ at $\ell$ positions, and in each step the allele $\alpha_i$
at some position $i$ is replaced by the allele $\omega_i$. This implies that the path is confined to a binary hypercube of
dimension $\ell$ \cite{Zagorski2016}, and the properties of the allele graph ${\mathcal{A}}$ are irrelevant (apart from
the fact that all mutations on the path have to be allowed). Because there are $\ell!$ paths in total, the
expected number of direct accessible paths
is
\begin{equation}
  \label{eq:expdir}
  \mathbb{E}^{\mathrm{dir}}[X_{\alpha,\omega}] = \ell! P_{\beta,\ell} = \ell
    \beta^{\ell-1},
  \end{equation}
 which vanishes for $\ell \to \infty$ whenever $\beta < 1$. By
 Markov's inequality
 \begin{equation}
   \label{eq:Markov}
   \mathbb{P}[X_{\alpha,\omega} \geq 1] = \sum_{k = 1}^\infty   \mathbb{P}[X_{\alpha,\omega} = k] \leq  \sum_{k = 1}^\infty k
   \mathbb{P}[X_{\alpha,\omega} = k] = \mathbb{E}[X_{\alpha,\omega}]
 \end{equation}
 we conclude that long direct accessible paths are possible only at
 the maximal fitness difference $\beta = 1$. A more refined analysis \cite{Hegarty2014}
 shows that the transition to high accessibility occurs
 at an $\ell$-dependent threshold
 \begin{equation}
   \label{eq:betaell}
   \beta_c(\ell) = 1 - \frac{\ln \ell}{\ell}
 \end{equation}
 in the sense that, for a sequence $\{\beta_\ell\}$ of $\ell$-dependent fitness differences,
 \begin{equation}
  \label{eq:HegartyMartinsson}
  \lim_{\ell \to \infty} \mathbb{P}[X_{\alpha,
    \omega} \geq 1 \vert \beta = \beta_\ell] = 
\begin{cases} \displaystyle 0 & \beta_\ell <  \beta_c(\ell)  \\
\displaystyle 1 & \beta_\ell >  \beta_c(\ell).
\end{cases}
 \end{equation}
 We note for later reference that the threshold (\ref{eq:betaell}) can be read off from the expected number
 of paths (\ref{eq:expdir}) by demanding that 
 \begin{equation}
   \label{eq:acccond}
 \lim_{\ell \to \infty}  \mathbb{E}^\mathrm{dir}[X_{\alpha,\omega} \vert \beta_\ell = \beta_c(\ell)] = 1.
 \end{equation}
 The full distribution of the number of direct accessible paths was found in \cite{Berestycki2016}. 

 Instead of constraining the fitness difference between the endpoints
 of the path, one may also condition the final point to be of high
 fitness $g(\omega) = 1$, and leave the fitness of the starting point $\alpha$
 undetermined \cite{Carneiro2010}. Averaging (\ref{eq:expdir}) over the uniformly
 distributed fitness of the starting point, one obtains 
 \begin{equation}
   \label{eq:dirpathsmax}
   \overline{\mathbb{E}^\mathrm{dir}[X_{\alpha,\omega}]} = \int_0^1 d
   \beta \;
   \ell \beta^{\ell-1}  = 1
 \end{equation}
independent of the path length $\ell$. A combinatorial derivation of
this result was given in \cite{Franke2011}.

 \subsection{Indirect paths}
 \label{sec:undirected}

  An indirect path is longer than the distance between its endpoints, $\ell > d(\alpha,\omega)$, which reduces the
 probability (\ref{eq:Pbetaell}) for such a path to be accessible. This is however offset by an enormous increase in the number
 of paths \cite{DePristo2007,Palmer2015,Wu2016,Zagorski2016}.
 The total number of self-avoiding paths on the $L$-cube grows double-exponentially in $L$ and is
 known explicitly only up to $L=5$ \cite{Berestycki2017}. As will be shown in this section, the enhancement of accessibility
 by indirect paths leads to the emergence of a nontrivial accessibility threshold $\beta_c \in (0,1)$. 

For general allele graphs ${\cal{A}}$ and $a > 2$, the mutual accessibility of two genotypes along general (indirect) paths
 depends not only on their distance, but also on their allelic composition. Asymptotically for large path length $\ell \to \infty$,
 the relevant properties of the end points $\alpha, \omega$ are encoded in the \textit{divergence matrix}
 $\mathbf{P} = \{p_{\mu \nu}\}_{\mu,\nu = 0,\dots,a-1}$,
 where $p_{\mu \nu}$ is the fraction of sites $i \in \{1,\dots,L\}$ at which $\alpha_i = \mu$ and $\omega_i = \nu$. The rescaled
 Hamming distance between the two genotypes is then given by
 \begin{equation}
   \label{eq:delta}
   \delta \equiv \lim_{L \to \infty} \frac{d(\alpha_L,\omega_L)}{L} = 1 - \sum_{\mu = 0}^{a-1} p_{\mu \mu}.
 \end{equation}
where $\{(\alpha_L, \omega_L) \in {\mathcal{A}}^L \times {\mathcal{A}}^L \}_{L \in \mathbb{N}}$ is a sequence of endpoints whose
asymptotic divergence matrix is given by $\mathbf{P}$.

 Similar to the direct path problem in section~\ref{sec:directed}, we approach the accessibility question through an analysis
 of the expected number of accessible paths. In principle, this amounts to enumerating all self-avoiding paths of a given
 length $\ell$, multiplying this number with the weight (\ref{eq:Pbetaell}), and summing over $\ell$ \cite{Berestycki2017}. 
 The difficulty of enumerating all self-avoiding paths can however be eliminated
 by introducing the notion of \textit{quasi-accessibility} \cite{Schmiegelt2023}. The accessibility
 problem is modified by allowing for multiple visits to the same genotype (thus removing the constraint of self-avoidance), but
 assigning a new i.i.d. fitness value to it at each visit. In the resulting \textit{extended HoC model} each genotype is thus
 equipped with a countable infinite sequence of fitness values which are evaluated in consecutive visits.
 A (generally self-intersecting) path on the genotype space ${\cal{A}}^L$ is then called quasi-accessible if the fitness values
 encountered in the extended HoC landscape are monotonically increasing. Denoting the number of quasi-accessible paths between
 genotypes $\alpha$ and $\omega$ by $\tilde{X}_{\alpha,\omega}$, it is proved in \cite{Schmiegelt2023} that
 \begin{equation}
   \label{eq:quasi}
   \mathbb{P}[\tilde{X}_{\alpha,\omega} \geq 1] = \mathbb{P}[X_{\alpha,\omega} \geq 1],
   \end{equation}
 and therefore the analysis of quasi-accessible paths is sufficient to determine the accessibility in the original model.

 A relatively straight-forward calculation shows that
 the exponential growth rate of the expected number of quasi-accessible paths is given by \cite{Schmiegelt2023}
\begin{equation}
 \label{eq:Gamma}
   \Gamma(\beta) \equiv  \lim_{L \to \infty} \frac{\ln \mathbb{E}[\tilde{X}_{\alpha_L,\omega_L}]}{L} =
   \sum_{\mu,\nu = 0}^{a-1} p_{\mu \nu} \ln \left[\left( e^{\beta \mathbf{A}} \right)_{\mu,\nu} \right],
\end{equation}
where $\left( e^{\beta \mathbf{A}} \right)_{\mu,\nu}$ is an element of the matrix exponential $e^{\beta \mathbf{A}}$.
Under certain natural conditions on the matrices $\mathbf{A}$ and $\mathbf{P}$, which are specified in \cite{Schmiegelt2023},
$\Gamma(\beta)$ is an increasing function with a unique zero, which we denote by $\beta^\ast$. For $\beta < \beta^\ast$
the function $\Gamma(\beta)$ is negative. Thus
the expected number of quasi-accessible paths tends to zero asymptotically, which implies by Markov's inequality
(\ref{eq:Markov}) that no quasi-accessible paths exists. By (\ref{eq:quasi}) this statement extends to the original
accessibility problem, and we conclude that $\beta^\ast$ is a lower bound on the accessibility percolation threshold
$\beta_c$. In particular, if $\beta^\ast > 1$, accessible paths between the two endpoint genotypes cannot exist for any
fitness difference $\beta \in (0,1]$. An example for which this can be shown to be
  the case is the path graph over $a \geq 3$ alleles
  [see figure \ref{fig:allelegraphs}c)].

  Determining whether the bound obtained from Markov's inequality is tight requires a careful analysis of higher moments
  of $\tilde{X}_{\alpha \omega}$ \cite{Schmiegelt2023}. The result of this analysis is a negativity condition on a certain function
  constructed from $\mathbf{A}$ and $\mathbf{P}$,
  which was previously introduced in the context of first passage percolation \cite{Martinsson2018}. Accessibility
  problems that satisfy this condition are said to be of \textit{regular type} \cite{Schmiegelt2023}. An example
  of an allele graph for which the negativity condition is violated, and for which the true accessibility threshold $\beta_c$ provably
  exceeds the lower bound $\beta^\ast <1$, is shown in figure~\ref{fig:allelegraphs}b). For problems of regular type 
  it is proved in \cite{Schmiegelt2023} that
  \begin{equation}
    \label{eq:C}
    \lim_{L \to \infty} \mathbb{P}[X_{\alpha_L,
        \omega_L} \geq 1 \vert \beta > \beta^\ast] \geq C
  \end{equation}
  for some positive constant $C$. Based on related results in \cite{Martinsson2018} and the numerical simulations
  reported in \cite{Zagorski2016}, it is conjectured in
  \cite{Schmiegelt2023} that the bound can actually be improved to its maximal value $C = 1$. Assuming this to be the case,
  we can summarize the percolation behavior for accessibility problems of regular type in the form
  \begin{equation}
  \label{eq:percolation}
  \lim_{L \to \infty} \mathbb{P}[X_{\alpha_L,
    \omega_L} \geq 1 \vert \beta] = 
\begin{cases} \displaystyle 0 & \beta <  \beta_c  \\
\displaystyle 1 & \beta >  \beta_c,
\end{cases}
  \end{equation}
where $\beta_c$ is the solution of $\Gamma(\beta_c) = 0$. Moreover,
the typical length $\ell^\ast$ of accessible paths near the threshold is given to
leading order in $L$ by
\begin{equation}
  \label{eq:pathlength}
  \lim_{L \to \infty} \frac{\ell^\ast(\alpha_L, \omega_L)}{L} =
  \beta_c \Gamma'(\beta_c)
\end{equation}
where $\Gamma' = \frac{d \Gamma}{d \beta}$ \cite{Schmiegelt2023}.

\subsection{Accessibility of the Hamming graphs}
\label{sec:Hamminggraph}

We recall that the Hamming graph $\mathbb{H}_a^L$ is the $L$-fold
Cartesian product of the complete allele graph over $a$ alleles.  
In this case the accessibility problem is of regular
type, and it is fully specified by the scaled Hamming distance
$\delta$ between the two endpoints. The exponential growth rate
(\ref{eq:Gamma}) of the
expected number of quasi-accessible paths is then given by
\begin{equation}
  \label{eq:Gammacomplete}
  \Gamma(\beta) = - \ln a - \beta + \delta \ln \left(e^{a\beta} - 1
  \right) + (1-\delta) \ln \left(e^{a\beta} + a - 1 \right).
  \end{equation}
For the binary case $a=2$ the threshold condition $\Gamma(\beta_c) =
0$ takes on the form
\begin{equation}
  \label{eq:binary}
  \sinh(\beta_c)^\delta \cosh(\beta_c)^{1-\delta} = 1,
  \end{equation}
which was first conjectured in \cite{Berestycki2017}, and
subsequently proved in \cite{Martinsson2015,Li2018}. For endpoints at
maximal distance ($\delta = 1$), $e^{\beta_c}$ is the solution of an
algebraic equation of order $a$
\begin{equation}
  \label{eq:algebraic}
  \left(\e^{\beta_c}\right)^a - a \e^{\beta_c} - 1 = 0,
  \end{equation}
which can be solved in closed form for $a \leq 4$ (see table \ref{Table1}). For large $a$ the
asymptotic expansions
\begin{equation}
  \label{eq:alarge1}
    \beta_c = \frac{\ln a}{a} + \frac{1+\ln a}{a^2} + {\mathcal{O}}\left(\frac{\ln
        a}{a^3}\right)
  \end{equation}
  and
  \begin{equation}
    \label{eq:alarge2}
   \beta_c \Gamma'(\beta_c)  = \ln a + \frac{1+\ln a}{a} +
   \mathcal{O}\left(\frac{\ln a}{a^2}\right)
 \end{equation}
 can be derived. Two general conclusions can be drawn from these
 expressions. First, HoC landscapes become much more accessible as the number
 of alleles increases, in the sense that $\beta_c \to
 0$ for $a \to \infty$. Second, the increase of accessibility is
 achieved by paths that are only moderately longer (by a factor $\sim
 \ln a$) than the direct paths.

\begin{center}
  \begin{table}
    \begin{tabular}{l|l|l|l}
        $a$ & $\beta_c$ & $\Gamma'(\beta_c)$ & $\beta_c \Gamma'(\beta_c)$ \\ \hline
        $2$ & $\arcsin(1) \approx 0.881$ & $\sqrt{2} \approx 1.41$ & $\approx 1.25$ \\
        $3$ & $\ln\left(2\cos\frac{\pi}{9}\right) \approx 0.631$ & $1+2\cos\left(\frac{2\pi}{9}\right) \approx 2.53$ & $\approx 1.82$ \\
      $4$ & $\ln\left(\frac{1}{\sqrt{2}}\right)+\sqrt{\sqrt{2}-\frac{1}{2}} \approx 0.509$ & $\approx 3.60$ & $\approx 1.83$
    \end{tabular}
    \caption{Results for the complete allele graph with 2-4 alleles
      and endpoints at maximal distance $\delta = 1$.
      The last column shows the scaled asymptotic path
      length (\ref{eq:pathlength}) at the threshold. In the biallelic case $a = 2$
      the result for the path length was obtained independently in
      \cite{Kistler2020} for a related model. This work also contains additional results on the structure
      of accessible paths. \label{Table1}}
    \end{table}
\end{center}
 
 The results
 presented here and in \cite{Schmiegelt2023} are in excellent agreement with earlier
 numerical simulations of accessible paths on Hamming graphs \cite{Zagorski2016}. The authors of \cite{Zagorski2016} focused
 on the probability of existence of at least one accessible maximal distance path ($d(\alpha,\omega) = L$)
 to the global fitness maximum. In this case the fitness of the endpoint genotype
 is effectively $g(\omega) = 1$, whereas the starting point has uniformly distributed fitness. The probability
 of existence of such a path is then given by $1-\beta_c$ and tends to
 unity for $a \to \infty$. For later reference we note that, because the fitness difference
$\beta \in [-1,1]$ between two arbitrary genotypes $\alpha,
 \omega$ has probability density $1 - \vert \beta \vert$, the probability for
 the existence of an accessible path $\alpha \to \omega$ between two randomly chosen genotypes at scaled distance $\delta$
 is asymptotically given by 
\begin{equation}
  \label{eq:pran}
  p_\mathrm{rand}(\delta) = \int_{-1}^1 d\beta \,(1 - \vert \beta \vert) \mathbb{P}[X_{\alpha,
      \omega} \geq 1 \vert \beta,
  \delta] = \int_{\beta_c(\delta)}^1 d\beta (1 - \beta)
  = \frac{1}{2} [1 - \beta_c(\delta)]^2.
\end{equation}

\section{Structured fitness landscapes}
\label{Sec:structured}

There is ample empirical evidence that real fitness landscapes are rugged, but they are not completely random
\cite{deVisser2014,Fragata2019,Bank2022}.
Quantitative measures of ruggedness, such as the fraction of local peaks among genotype sequences, generally fall
in between the limits of uncorrelated random and smooth (single-peaked) landscapes \cite{Aguilar2017,Szendro2013}. In this
section we examine the consequences that different kinds of landscape structure have on evolutionary accessibility.
While it might be expected that reduced ruggedness implies increased accessibility, we will see that this is not
generally true.

\subsection{Geometry of the binary hypercube}

\label{Sec:binary}

Since theoretical work on models of structured fitness landscapes is largely restricted to the case of binary
sequences, throughout this section we take the sequence space to be the binary hypercube $\mathbb{H}_2^L$. To
conveniently describe the geometry of this space, we introduce additional concepts that are specific
to the binary case. First, the number
\begin{equation}
  \label{eq:nsigma}
  n_\sigma = \sum_{i=1}^L \sigma_i
\end{equation}
of 1's in the sequence $\sigma$ provides a natural one-dimensional coordinate that runs from $n_\sigma = 0$
for the 0-string $\sigma = (0,\dots,0)$ to $n_\sigma = L$ for the 1-string $\sigma = (1,\dots,1)$. Referring to the 0-string
as the \textit{wild type}, $n_\sigma$ counts the number of mutations relative to the wild type, and
the 1-string is the \textit{full mutant}. A path in $\mathbb{H}_2^L$ is called \textit{mutation directed} (or \textit{directed}
for short) if $n_\sigma$
changes monotonically along the path.

Second, it is often useful to represent a genotype $\sigma$ by the subset of sites $i$ in the \textit{locus set}
${\mathcal{L}} = \{1,\dots,L\}$ at which $\sigma_i = 1$ \cite{Das2020}. In this notation, the wild type is
the empty set $\emptyset$, and the full mutant is identified with ${\mathcal{L}}$. An immediate corollary is that a
mutation directed
path from $\alpha$ to $\omega$ exists iff $\alpha$ is a subset of $\omega$ or vice versa. Such a path
is then also direct in the sense of section \ref{Sec:Space}. Finally,
the \textit{antipode} $\bar{\sigma}$ of a binary sequence is obtained
by reverting all the entries,
\begin{equation}
  \label{eq:antipode}
  \bar{\sigma_i} = 1 - \sigma_i.
\end{equation}
The antipode is the unique sequence at maximal distance $L$ from $\sigma$.

\subsection{Additive landscapes}
\label{sec:additive}

It is instructive to begin the discussion of structured landscapes with the seemingly trivial case of an
additive landscape with a single peak \cite{Schmiegelt2023a}. The fitness function is of the form
\begin{equation}
  \label{eq:additive}
  g(\sigma) = \sum_{i=1}^L a_i \sigma_i,
\end{equation}
and we assume for now that the coefficients $a_i$ are positive. Then the full mutant is the global and unique fitness peak, and
all mutation directed paths to the peak are accessible. For a pair of genotypes $\alpha, \omega$, a directed path
$\alpha \to \omega$ exists if
\begin{equation}
  \label{eq:subset}
  \alpha \subset \omega,
\end{equation}
i.e., $\alpha$ is a subset of $\omega$ or (equivalently) $\omega$ is a \textit{superset} of $\alpha$. 
Thus the peak is accessible from all genotypes ($\omega = {\mathcal{L}}$) and all genotypes can be accessed from
the wild type ($\alpha = \emptyset$). 

For two randomly chosen genotypes $\alpha, \omega$ at Hamming distance $d$,
the subset condition (\ref{eq:subset}) requires that for all $d$ sites
at which the two sequences differ the corresponding entries are given by $\alpha_i = 0$ and $\omega_i = 1$. This is true with
probability $p_\mathrm{rand}(d) = 2^{-d}$, which [in contrast to the corresponding quantity (\ref{eq:pran}) in the random
  case] tends to zero for $d \to \infty$. If (\ref{eq:subset}) holds there are $d!$ different direct and directed
paths from $\alpha$ to $\omega$, and thus the expected number of accessible paths between random pairs
of genotypes is given by
\begin{equation}
  \label{eq:expadditive}
  \mathbb{E}^\mathrm{add}[X_{\alpha, \beta} \vert d] = 2^{-d} d!
\end{equation}
which grows combinatorially for large $d$. 

These statements remain valid if the coefficients $a_i$ in (\ref{eq:additive}) are random numbers without a definite sign,
except that the unique and global peak is then given by the sequence with entries $\sigma_i = \frac{1}{2}[1 +
  \mathrm{sgn}(a_i)]$.
The previous situation is recovered through a symmetry operation of the hypercube
that maps this sequence to the full mutant. We conclude that
for additive landscapes accessibility between arbitrary pairs of genotypes is low, but there is a direction of high
accessibility, which generally depends on the $a_i$. Of course the low
accessibility between random pairs is no obstacle for the evolutionary dynamics, since any strategy that accepts beneficial mutations and rejects
deleterious ones will carry the population to the global fitness peak.

\subsection{Kauffman's NK-model}

The NK-model provides a simple framework for generating instances of
fitness landscapes with a tunable degree of ruggedness
\cite{Kauffman1989,Weinberger1991}. The basic idea is to write the fitness as a sum
of contributions, each of which depends randomly on the variables at a
subset of the locus set $\mathcal{L}$. The main tuning parameter of
the model is the size of the subsets, which is generally assumed to be
the same for all subsets and will be denoted here by $k$. When the
subsets consist of single loci, $k=1$, the model reduces to the
additive landscape (\ref{eq:additive}) with random coefficients,
whereas for $k=L$ the fitness is a random function of the entire
sequence, and one recovers the HoC model discussed in section
\ref{Sec:random}.

\begin{figure}[htb]
    \centering
    \includegraphics[width=0.3\textwidth]{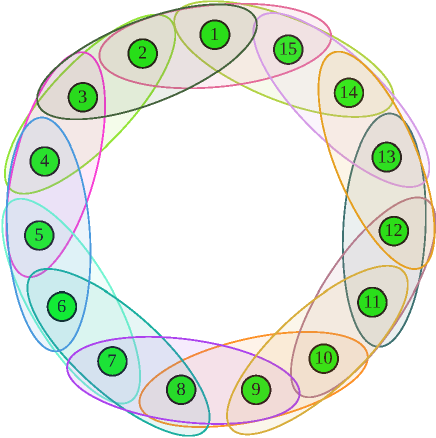}\hspace*{1.cm}\includegraphics[width=0.3\textwidth]{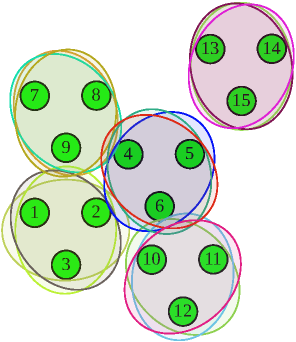}\hspace*{1.cm}\includegraphics[width=0.3\textwidth]{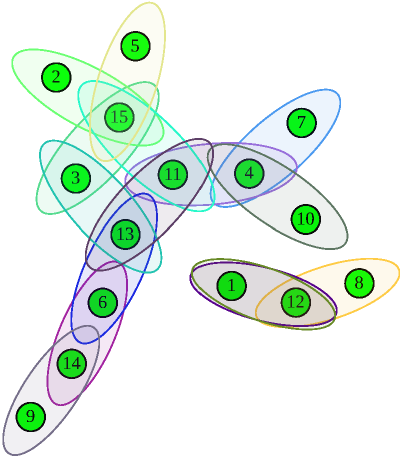}
    \caption{\textbf{Interaction structures in the NK-model.} The
      figure illustrates possible choices of interacting subsets
      among $L=15$ loci. The loci enclosed in an ellipse belong to the same interaction set $B_i$.
      Note that the sets are not necessarily distinct. In the left and middle panel the subsets have
    size $k=3$, and in the right panel $k=2$. In the left panel the
    loci are arranged on a ring (\textit{adjacent structure}), in the
    middle panel the subsets are either disjoint or identical (\textit{block structure})
    and in the right panel the subsets are chosen randomly
    (\textit{random structure}). Modified from \cite{Hwang2018}.}
    \label{fig:NK}
\end{figure}

For a formal definition, we introduce the \textit{interaction structure} \cite{Hwang2018}
of the model as a collection of subsets $B_i \subset {\mathcal{L}}$ of
size $\vert B_i \vert = k$, where $i = 1, \dots, b$, and write the
fitness function as
\begin{equation}
  \label{eq:NKfitness}
  g(\sigma) = \sum_{i=1}^b \phi_i^{(k)}(\downarrow_{B_i} \!\! \sigma).
  \end{equation}
The map $\downarrow_{\mathcal{S}}: \mathbb{H}_2^L \to
\mathbb{H}_2^{\vert {\mathcal{S}} \vert}$ projects the sequence
  $\sigma$ onto the subspace spanned by the loci in the subset
  ${\mathcal{S}}$, and the $\phi_i^{(k)}$ are independent realizations of HoC fitness
  landscapes on $\mathbb{H}_2^k$. Examples of commonly used
  interaction structures are shown in figure \ref{fig:NK}.

The analysis of the NK-model is straightforward for the case of
the block structure (middle panel of  figure \ref{fig:NK})
\cite{Perelson1995,Schmiegelt2014}. We assume that $L/k$ is an integer, and partition the locus set into
$b = L/k$ disjoint interaction sets $B_i$, which will be referred to as blocks.
Consider first the number of 
fitness peaks, which we denote by $N$. A sequence $\sigma$ is a
local fitness maximum iff each of the projected sequences
$\downarrow_{B_i} \!\!\! \sigma, i = 1,\dots,b$ is a maximum in its
respective sublandscape. Denoting the number of peaks of the $i$-th
sublandscape by $N_i$, we thus have
\begin{equation}
  \label{eq:NKpeaks1}
  N = \prod_{i=1}^{L/k} N_i.
\end{equation}
Since the sublandscapes are independent, we can average over both
sides to obtain
\begin{equation}
  \label{eq:NKpeaks2}
  \mathbb{E}[N] = [\mathbb{E}(N_i)]^{L/k} =
  \left[\frac{2^{k}}{k+1}\right]^{L/k},
\end{equation}
where we have used that the mean number of peaks in a HoC
landscape on $\mathbb{H}_a^L$ is \cite{Kauffman1987}
\begin{equation}
  \label{eq:HoCpeaks}
  \mathbb{E}^\mathrm{HoC}[N] = \frac{a^L}{(a-1)L+1}.
\end{equation}
  As expected, the expression (\ref{eq:NKpeaks2})
interpolates monotonically between the minimally rugged additive case ($k=1$)
and the maximally rugged HoC landscape ($k=L$).
The computation of the mean number of peaks for general
interaction structures is nontrivial, but partial results
are available and have been reviewed in \cite{Hwang2018}. Remarkably,
the asymptotic behavior of $\mathbb{E}[N]$ for large $L$ is generally
quite close to that predicted by the simple expression
(\ref{eq:NKpeaks2}) for the block structure, and (\ref{eq:NKpeaks2}) has been
conjectured to constitute an upper bound for general interaction
structures \cite{Schmiegelt2014}.

A similar approach can be used to examine the evolutionary
accessibility of the NK-model with block interaction structure
\cite{Schmiegelt2014}. For simplicity we focus on direct paths of
length $L$ that end at the global fitness maximum (henceforth denoted by $\omega$) and
start at its antipode $\alpha = \bar{\omega}$. The projection
$\downarrow_{B_i} \! \omega$ of the global maximum to the $i$-th block
is the global maximum of the corresponding sublandscape, and each step
along the path has to increase the fitness contribution $\phi_i^{(k)}$ of
the block in which the corresponding locus resides. Thus, each direct
accessible path $\alpha \to \omega$ can be decomposed into accessible
direct subpaths $\downarrow_{B_i} \! \alpha \to \downarrow_{B_i} \!
\omega$. Whereas the ordering of the steps within a sublandscape is
fixed by the subpath, steps in different sublandscapes
can occur in arbitrary order. Taken together, these considerations
imply that the number of direct accessible paths can be written as  
\begin{equation}
  \label{eq:blockpaths}
  X_{\alpha,\omega} = \frac{L!}{(k!)^b} \prod_{i=1}^b
    X_{\downarrow_{B_i} \alpha,\downarrow_{B_i} \omega},
  \end{equation}
  where the combinatorial prefactor accounts for the different ways in
  which a given set of subpaths can be combined. Averaging both sides
  and making use of the result (\ref{eq:dirpathsmax}) for the expected
  number of direct paths to the global maximum of an HoC landscape, we
  obtain
  \begin{equation}
    \label{eq:blockpathsav}
    \mathbb{E}[X_{\alpha,\omega}] = \frac{L!}{(k!)^b},
  \end{equation}
  which again encompasses the limiting cases $k=1$ and $k=L$, and
  grows combinatorially fast with $L$ for any fixed $k$.

  On the other
  hand, for the global maximum of the full landscape to be at all
  accessible from the antipode, at least one accessible path has to
  exist in every sublandscape. This implies that
  \begin{equation}
    \label{eq:blockpathexist}
    \mathbb{P}[X_{\alpha,\omega} \geq 1] =
    \mathbb{P}[X_{\downarrow_{B_i} \alpha,\downarrow_{B_i} \omega}
    \geq 1]^{L/k}.
  \end{equation}
  Since $ \mathbb{P}[X_{\downarrow_{B_i} \alpha,\downarrow_{B_i}
    \omega} \geq 1] < 1$ for any $k > 1$, this quantity decays exponentially
  with $L$. Note that the exponential
  decay is much faster than the corresponding behavior for direct paths in the
  random HoC model considered in section \ref{sec:directed}, where (when conditioning the endpoint to be of
  maximal fitness)
  \begin{equation}
    \label{eq:NK_HoC}
   \mathbb{P}[X_{\alpha,\omega} \geq 1] = \mathbb{P}[\beta >
   \beta_c(L)] \approx \frac{\ln L}{L}.
 \end{equation}

  These considerations can be generalized to indirect paths \cite{Hwang2018}. The length
  $\ell$ of a path covering the maximal distance $L$ between the global fitness maximum and its antipode is a random
  variable, and similarly the number of (forward or backward) steps $\ell_i \geq k$ that the path spends in the $i$-th
  sublandscape is random, with $\sum_{i=1}^b \ell_i = \ell$. The multiplicity of each such path corresponding to the
  combinatorial prefactor in (\ref{eq:blockpaths}) is given by the multinomial coefficient
  \begin{equation}
    \label{eq:multinomial}
    \binom{\ell}{\ell_1, \dots, \ell_b} = \frac{\ell!}{\prod_{i=1}^b \ell_i!}.
  \end{equation}
  Importantly, (\ref{eq:blockpathexist}) remains valid and shows that the probability for existence of an accessible path decays
  exponentially in $L$ for any fixed $k$. The results described in section \ref{sec:Hamminggraph} show that, for large
  $k$, $\mathbb{P}[X_{\downarrow_{B_i} \alpha,\downarrow_{B_i} \omega}]$ converges to $1-\beta_c(\delta = 1, a=2) \approx 0.119$. A nonzero limit
  of (\ref{eq:blockpathexist}) is possible only if the number of interaction sets remains finite,
  which implies that $k$ increases proportionally to $L$. 
  
 We conclude, therefore, that the accessibility of the global
  maximum in 
  the NK-landscape with block structure and fixed $k$ is low, and remarkably similar to the
  accessibility between random genotypes in the additive model
  discussed in section \ref{sec:additive}: The existence of paths is
  exponentially unlikely, but the expected number of paths grows
  combinatorially fast with path length. Refined arguments presented in \cite{Hwang2018} and \cite{Schmiegelt2016} show
  that the low accessibility of the block model is in fact typical of most common NK interaction structures, including the
  adjacent and random structures displayed in figure \ref{fig:NK}. The precise statement is that, for interaction structure graphs
  that satisfy a certain local boundedness condition,
  accessible paths with $d(\alpha,\omega) =  L$ are exponentially unlikely for
  $L \to \infty$. This behavior is strikingly different from that of uncorrelated random landscapes, where such paths exist
  with certainty provided the fitness difference between the endpoints is sufficiently large. 
  
  \subsection{Rough Mount Fuji landscapes}
  \label{Sec:RMF}

 The example of the NK-model shows, somewhat counterintuitively, that
 the reduced ruggedness of certain types of structured landscapes can
 decrease their accessibility. An example of a landscape that displays
 the expected negative correlation between ruggedness and accessibility is the
 Rough Mount Fuji (RMF) model defined as a weighted linear superposition of an additive and a
 random (HoC) landscape \cite{Aita2000,Neidhart2014},
 \begin{equation}
   \label{eq:RMF}
   g(\sigma) = \xi_\sigma + c d(\sigma,\sigma^\ast). 
   \end{equation}
   Here the $\xi_\sigma$ are continuous i.i.d. random variables, $c >
   0$ a constant and $\sigma^\ast$ a reference sequence. For $c=0$ the
   landscape is random, and when $c$ is sufficiently large (larger
   than the fluctuations of the $\xi_\sigma$) it becomes additive with
   a unique global peak at the antipode $\overline{\sigma^\ast}$ of the
   reference sequence. While the number of peaks decreases monotonically
   with increasing $c$ \cite{Neidhart2014}, the probability of
   existence of direct accessible paths from $\sigma^\ast$ to
   $\overline{\sigma^\ast}$ tends to unity for $L \to \infty$ for any $c >
   0$ \cite{Hegarty2014}.

  \section{Highly rugged yet highly accessible fitness landscapes}

  \label{Sec:highly}

 Empirical evidence suggests that real fitness landscapes may be both highly rugged
  and highly accessible. For example, a recent study of a fitness landscape comprising almost all $4^9 = 262,144$
  combinations of nucleotide mutations at 9 sites in the \textit{folA} 
  gene in \textit{E. coli} identified 514 peaks among the 18,019 sequences that gave rise to functional proteins \cite{Papkou2023}.
  To put this number into perspective, we note that, according to (\ref{eq:HoCpeaks}),
  the probability for a randomly chosen genotype in a HoC fitness landscape
  with $a$ alleles and sequence length $L$ to be a fitness peak is
  $\frac{1}{(a-1)L+1}$;
  this is simply the probability for the peak genotype to be the
  largest among $(a-1)L+1$ continuous i.i.d. random variables
   \cite{Kauffman1987}. Since most
  of the functional sequences of the \textit{folA} landscape have the
  maximal number of $(a-1)L = 27$ functional neighbors, a completely random landscape would have on average $18,019/28 \approx 644$ peaks, only slightly more than the
  experimental value. Thus, by any measure, the empirical landscape is highly rugged. However, at the same
  time the 73 highest fitness peaks (those whose fitness value exceeds the wild type fitness) were highly accessible,
  in the sense that their basins of attraction comprised a significant fraction of all genotypes, with a median value
  of 69 \%.

  Here and in the following the \textit{basin of attraction} (BoA) of a fitness peak consists of all genotypes from which
  the peak can be reached via an accessible path. Note that this definition differs from the \textit{gradient basins} considered
  in previous work, where each point in a landscape is uniquely assigned to the peak that is reached by a greedy (steepest
  ascent) walk starting from the point \cite{Wu2016,Weinberger1991,Wolfinger2004,Franke2012}.
  Whereas gradient basins partition the genotype
  space into disjoint sets, the BoA's considered here can (and do) overlap. Like accessible paths, they depend only
  on the orientation of the fitness graph induced by the landscape, and not on the actual fitness values. 

  In this section we describe a recently discovered class of fitness landscape models that combine a high degree of ruggedness
  with high accessibility. These landscapes are characterized by a certain accessibility property
  that we define in the next subsection. The accessibility property was first derived for a model of evolution in
  changing environments motivated by the evolution of antibiotic resistance \cite{Das2020,Das2022}, and this
  context is explained in section \ref{Sec:TIL}. In section \ref{Sec:universal} we provide a simple sufficient condition
  for the accessibility property,
  which allows us to devise a minimal representative for this class of models and to explore some of its
  properties (section \ref{Sec:intermediate}). Throughout we consider landscapes defined on the binary hypercube $\mathbb{H}_2^L$. The generalization of
  these concepts to multiallelic fitness landscapes with $a > 2$, which is required for applying them to the empirical landscape
  of \cite{Papkou2023} discussed above, remains a
  task for future research (see section \ref{Sec:Conclusions}).

  \begin{figure}[htb]
    \centering
    \includegraphics[width=0.8\textwidth]{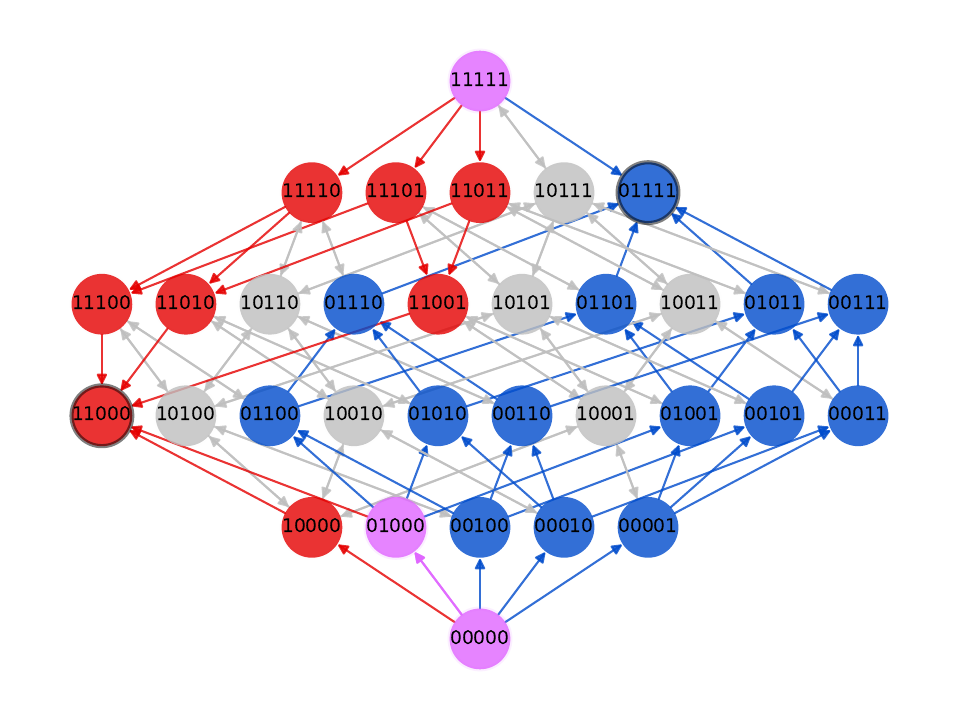}
    \caption{\textbf{Illustration of the subset-superset accessibility property.} The fitness graph of dimension $L=5$ has two peak genotypes,
      (0,1,1,1,1) and (1,1,0,0,0),
      which are highlighted by black circles. The direction of the colored arrows is forced by the AP, whereas the orientation of the
      light-grey arrows is arbitrary. Genotypes that are sub- or supersets of (0,1,1,1,1), and thus must belong to the corresponding BoA,
      are colored blue,  and the genotypes that must belong to the BoA of (1,1,0,0,0) are red. Purple genotypes are sub- or supersets of
      both peaks. 
    }
    \label{fig:APgraph}
\end{figure}

  \subsection{The accessibility property}
  \label{sec:AP}
  For the formulation of the property of interest we make use of the set
  notation introduced in section \ref{Sec:binary}. Importantly, this implies that the landscapes we consider are
  \textit{graded}, in the sense that their properties vary systematically along the coordinate $n_\sigma$ connecting the wildtype
  to the full mutant.
  In this respect they differ from the HoC and NK-models discussed
  previously, which are statistically isotropic (averaged over the
  random variables defining the landscape). By contrast, the RMF landscapes considered in section \ref{Sec:RMF} are similarly
  structured along the coordinate defined by the distance to the reference sequence. 

  We say that a fitness landscape has the \textit{subset-superset accessibility property (AP)} if the following holds \cite{Das2020}: 

  \begin{itemize}

  \item[] \textit{Any peak genotype is accessible from all its subsets and supersets via all direct paths.}

  \end{itemize}
  This implies that the basins of attraction of fitness peaks have a simple structure that is reminiscent of the additive
  landscapes of section \ref{sec:additive}. The BoA of a peak genotype $\sigma$ with $n_\sigma$ mutations
  consists of the two subcubes spanned by $\sigma$ and $\emptyset$ (the subsets of $\sigma$) and by $\sigma$ and
  ${\mathcal{L}}$ (the supersets of $\sigma$), respectively. All $n_\sigma!$ (direct and directed) paths 
  $\emptyset \to \sigma$
  are accessible, as well as all $(L-n_\sigma)!$ paths ${\mathcal{L}} \to \sigma$, and the size $S_\sigma$ of the
  BoA is bounded from below by
  \begin{equation}
    \label{eq:basin}
    S_\sigma \geq 2^{n_\sigma} + 2^{L-n_\sigma} - 2.
  \end{equation}
  Here we adopt the convention that a genotype does not belong to its own BoA.
  Note that by definition $S_\sigma \geq L$. An example of a fitness graph with two peaks
  that satisfies the AP is shown in figure \ref{fig:APgraph}.

  Next we derive an upper bound on the number of peaks $N$ that a landscape satisfying the AP (an AP-landscape for short)
  can have \cite{Oros2022}.
  Since no peak can be in the basin of attraction of another peak, peaks cannot be subsets or supersets of each other.
  According to Sperner's theorem, the largest collection of subsets of a set of $S$ elements that has this property is of
  size ${S \choose {\lfloor S/2 \rfloor}}$ \cite{Sperner1928}, and we conclude that
  \begin{equation}
    \label{eq:APpeaks}
    N \leq {L \choose {\lfloor L/2 \rfloor}}.
  \end{equation}
  The bound is saturated by a fitness landscape where all genotypes with $n_\sigma = \lfloor L/2 \rfloor$ are peaks
  and no other peaks exist. For large $L$ the bound (\ref{eq:APpeaks}) is of order $\frac{2^L}{\sqrt{\pi L/2}}$, which
  differs from the maximal number of peaks in an unconstrained landscape, $2^{L-1}$ \cite{Haldane1931,Crona2023}, only by a polynomial
  factor.

  \subsection{Tradeoff-induced fitness landscapes}
\label{Sec:TIL}

  The model of
tradeoff-induced fitness landscapes (TIL-model) describes evolutionary
processes of a population facing a tradeoff in its adaptation to
two extreme environments \cite{Das2020,Das2022}. The model generates a family of fitness
landscapes parametrized by a single, continuous environmental parameter. The scenario guiding the construction of the
model is the evolution of antibiotic resistance in bacteria, where the
environmental parameter determining the strength and direction of
selection is the drug concentration. In this context, the \textit{dose-response
  curve} $g(\sigma,c)$ represents the growth rate of a bacterial population with genotype $\sigma$ exposed
to the drug concentration $c$ \cite{Regoes2004}. Motivated by empirical
findings laid out in \cite{Das2020}, the dose-response curves are
assumed to be of the form
\begin{equation}
  \label{eq:TIL}
  g(\sigma,c) = r_\sigma f(c/m_\sigma).
\end{equation}
 The function $f(x)$ is monotonically
decreasing, accounting for the suppression of bacterial growth with
increasing drug concentration. Setting $f(0) = 1$, the parameter
$r_\sigma$ is identified as the \textit{drug-free growth rate} of the
strain, whereas $m_\sigma$ defines the concentration scale at which
the growth rate is significantly suppressed. In other words,
$m_\sigma$ quantifies the \textit{resistance} of the strain. 
With increasing $x$ the function $f(x)$ may either reach zero growth rate at
some finite value $x_c$, in which case $c = m_\sigma x_c$ is called the
minimal inhibitory concentration (MIC), or $f(x) \to 0$ asymptotically
for large $x$. A commonly used functional form of the latter type is
the Hill function
\begin{equation}
  \label{eq:Hill}
  f(x) = \frac{1}{1+x^h}
  \end{equation}
with a Hill coefficient $h > 0$. 

It follows from (\ref{eq:TIL}) and (\ref{eq:Hill}) that strains with large $r_\sigma$ are
favored at low concentrations, whereas strains with large $m_\sigma$
have an advantage at high concentrations. A tradeoff between these two
extremes occurs if strains with large $r_\sigma$ have small $m_\sigma$
and vice versa. A simple (and empirically supported) way of
implementing such a tradeoff is to write the drug-free growth rate and
the resistance in the form
\begin{equation}
  \label{eq:tradeoff}
  r_\sigma = \prod_{i=1}^L r_i^{\sigma_i}, \;\;\;\;\; m_\sigma =
  \prod_{i=1}^L m_i^{\sigma_i},
\end{equation}
where $r_i < 1$ and $m_i > 1$ are parameters associated with the $i$'th
mutation that can be measured experimentally or
drawn from some suitable probability distribution \cite{Das2020}.  
Equation (\ref{eq:tradeoff}) implies that genotypes with many
mutations (large $n_\sigma$) have, on average, low drug-free growth
rate and high resistance.

As a consequence of
the tradeoff between drug-free growth rate and resistance
the dose-response curves of different strains may cross \cite{Das2020}. If
the crossing occurs between strains whose genotypes are mutational
neighbors, the corresponding arrow in the fitness graph flips.
Under suitable (mild) conditions on the scaling function $f(x)$ and the parameters $r_i$ and $m_i$, it can be ensured that
the dose-response curves of neighboring genotypes cross exactly once \cite{Das2020}. 
In this
way the TIL-model defined by equations (\ref{eq:TIL}-\ref{eq:tradeoff})
induces an evolution on the set of fitness graphs which is
parametrized by $c$.

\begin{figure}[htb]
    \centering
     \includegraphics[width=0.7\textwidth]{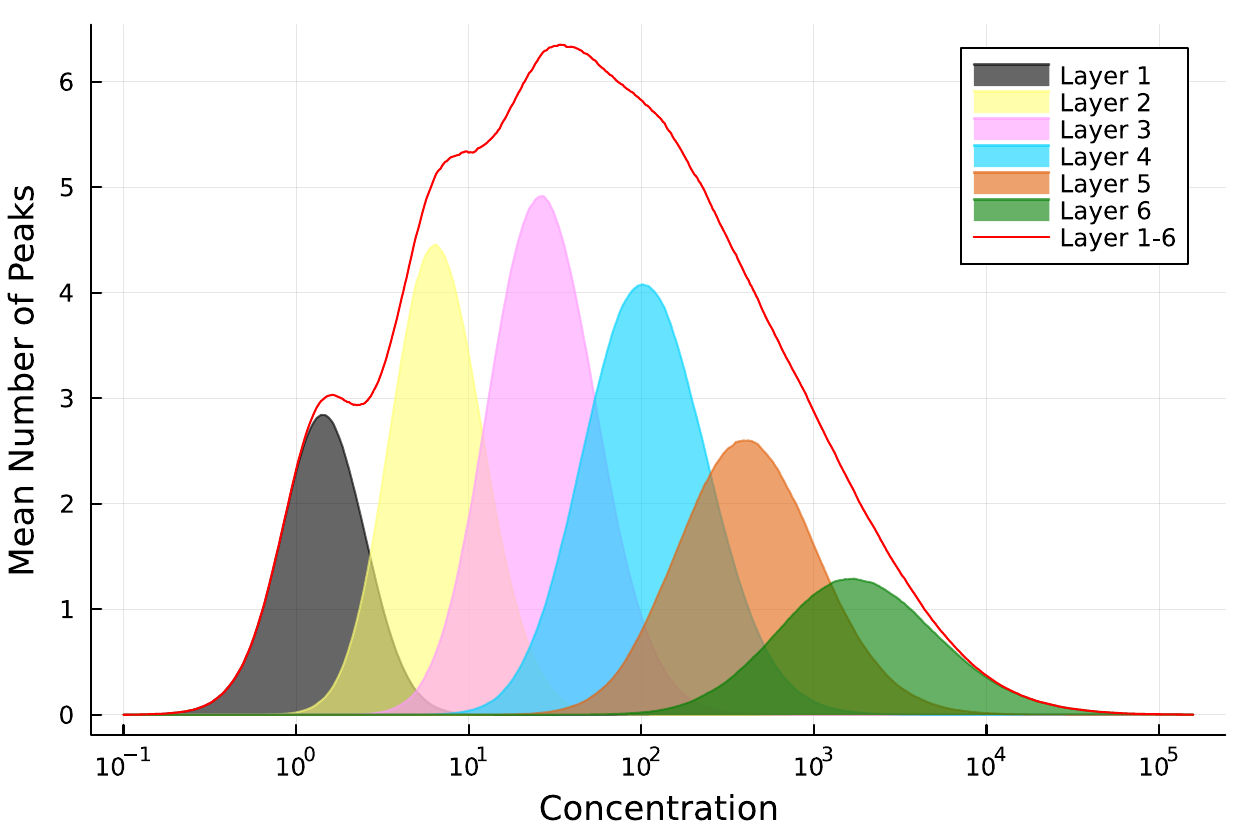}
     \caption{\textbf{Fitness maxima in tradeoff-induced landscapes.} The figure shows the mean number of fitness maxima
       in the TIL model as a function of concentration $c$, obtained from 15000 realizations with
      $L=7$. The parameters $\{r_i\}_{i=1,\dots,L}$ and $\{m_i\}_{i=1,\dots,L}$ were sampled from the joint
        distribution described in \cite{Das2020} and the concentration
        was varied from $10^{-1}$ to $10^8$ in $1000$ logarithmically spaced steps. The shaded curves show the number of peaks
       in a particular \textit{layer} of sequence space, where the mutation
       number $n_\sigma$ is fixed, and the red line shows the total number of peaks in layer 1-6.
       At very small and very large concentrations the fitness landscape is single peaked with the unique fitness maximum 
       in layer $n_\sigma = 0$ and $n_\sigma = 7$, respectively. These peaks are not included in the figure. 
    }
    \label{fig:TIL_peaks}
\end{figure}

For small
and large $c$ the graphs are simple, because the corresponding fitness
landscapes are multiplicative:
\begin{equation}
  \label{eq:TIL-limits}
  g(\sigma, 0) = r_\sigma, \;\;\; g(\sigma, c \to \infty) \approx
  \frac{r_\sigma m_\sigma^h}{c^h} = c^{-h} \prod_{i=1}^L (r_i
  m_i^h)^{\sigma_i}.
\end{equation}
In these limits the fitness landscape is single peaked, and all
direct paths to the peak are accessible\footnote{If $g$ is
  multiplicative, $\ln g$ is
  additive in the sense of section \ref{sec:additive}.}. However, at
intermediate concentrations the tradeoff induces a significant amount
of ruggedness, and it can be shown that the number of peaks grows
exponentially with $L$ \cite{Das2020,Das2022}. The genotypes that
optimize the tradeoff (and therefore have highest fitness) 
at concentration $c$ have on the order of 
\begin{equation}
  \label{eq:nopt}
n_\mathrm{opt} \approx \frac{\ln c}{\langle \ln m_i \rangle}
  \end{equation}
mutations, and the ruggedness is maximal when $n_\mathrm{opt} \approx
\frac{L}{2}$ (figure \ref{fig:TIL_peaks}).
The accessibility property explained in section \ref{sec:AP} arises in
the TIL-model as a consequence of the ordering of the concentration values at which
the dose-response curves of neighboring genotypes cross, and it holds at any concentration (see
\cite{Das2020,Das2022} for details). 

\begin{figure}[htb]
    \centering
    \includegraphics[width=\textwidth]{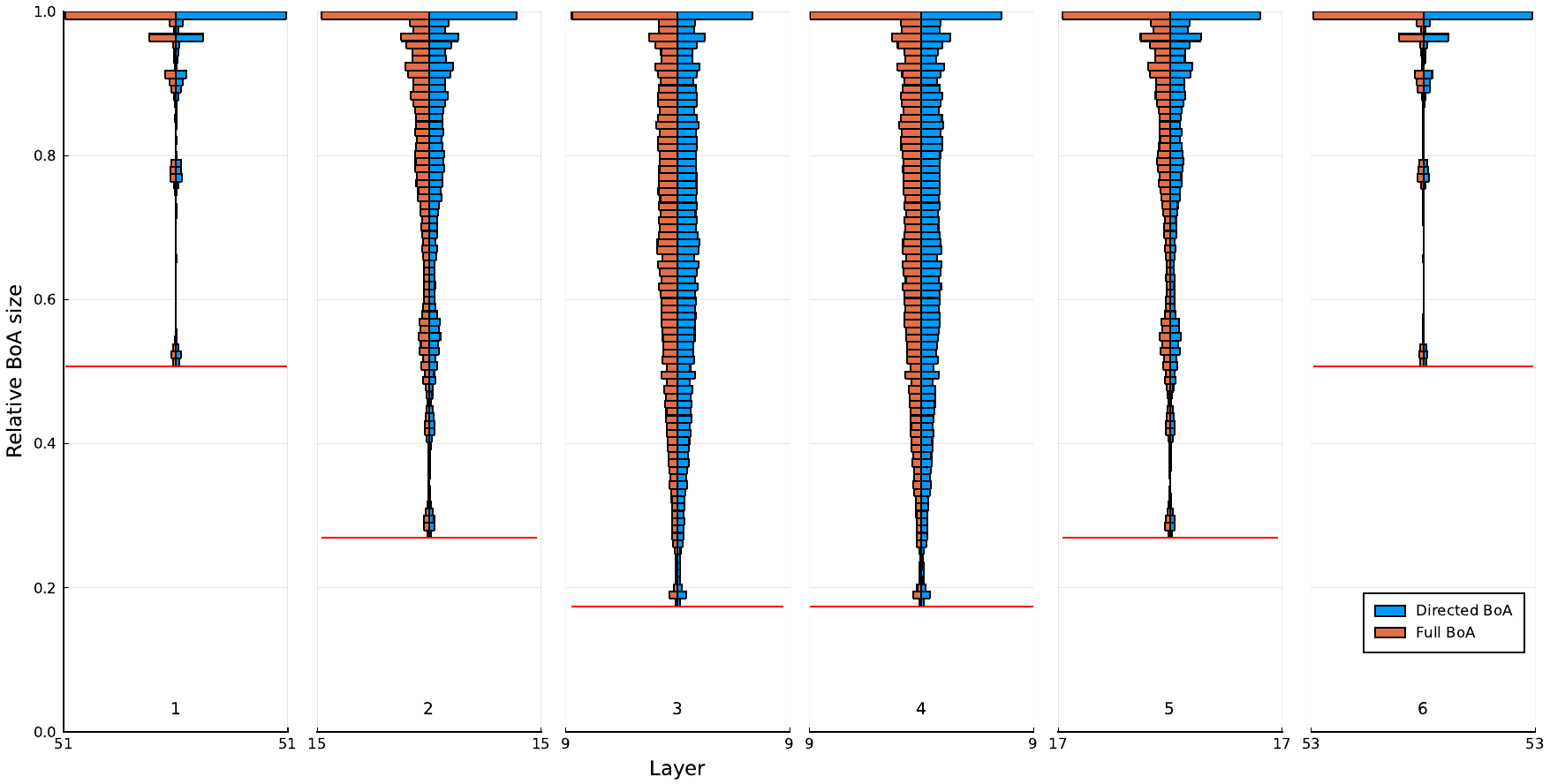}
    \caption{\textbf{Basins of attraction of fitness maxima in
        tradeoff-induced landscapes.} The figure shows histograms for
      the relative basin size (\ref{eq:BoA_rel})
      obtained from simulations of the TIL model with
      $L=7$ (see figure \ref{fig:TIL_peaks} for details).
      The peak genotypes are grouped along the $x$-axis into layers of equal mutation
      number. Only instances are shown where neither the
      wild type ($n_\sigma = 0$) nor the full mutant ($n_\sigma = 7$)
      are peaks, because in these cases the landscape must be single-peaked.
      For each peak genotype, the
        \textit{maximal} size of its BoA attained at any concentration
        is recorded.  In the vertical histograms, left orange bars refer to the relative size of the
      full BoA's spanned by all accessible paths, and right
      blue bars to the relative size of the BoA's spanned by directed
      paths. Note the different horizontal scale of each vertical
      histogram pair. The red horizontal lines mark the lower bound
      (\ref{eq:basin}) divided by $2^L - 2 =
      126$. This normalization accounts for the fact that, if a peak
      has a BoA of minimal size, there must be at least one other peak
      that accomodates the remaining genotypes. As a consequence, the
      maximal number of non-peak genotypes is $2^L-2$. 
    }
    \label{fig:TIL_BoA}
\end{figure}

In \cite{Das2020} the accessibility of the TIL landscape was
quantified through the reachability of the global fitness peak in
adaptive walk simulations, and it was found to be much higher than for
an NK-landscape of comparable ruggedness. Here we explore the accessibility of the landscape by measuring the
sizes of the BoA's of fitness peaks (figure \ref{fig:TIL_BoA}).
The relative basin size
$S_\sigma^\textrm{rel}$ depicted in the figure is obtained by 
normalizing the basin size $S_\sigma$ by the number of non-peak
genotypes,
\begin{equation}
  \label{eq:BoA_rel}
  S_\sigma^\textrm{rel} = \frac{S_\sigma}{2^L -N}.
\end{equation}
The figure shows that BoA's that are
considerably larger than the lower bound (\ref{eq:basin}) are common,
and in fact many have maximal size. Inspection of the data shows that most of
these instances are not single-peaked landscapes (for which trivially $S_\sigma^\textrm{rel} = 1$).
Interestingly, the histograms representing the
full and the directed BoA's have very similar shapes.

\subsection{Universal epistasis}

\label{Sec:universal}

\hspace*{0.5cm} \hfill \textit{``Epistasis'', like ``invertebrate'',
is a term that really means ``everything else''} \\
\hspace*{0.5cm} \hfill Sean H. Rice \cite{Rice2000}

In modern evolutionary genetics, the term \textit{epistasis} is used to
describe any kind of interaction in the effects of different
mutations on a phenotype or on fitness
\cite{deVisser2014,Bank2022,Weinreich2005,Wolf2000,Phillips2008,Ferretti2016,Domingo2019,Krug2021a}. In the present context this
means that the fitness function $g(\sigma)$ deviates from the additive
form (\ref{eq:additive}).
As an illustrative example, consider two mutations occurring at
positions $i$ and $j$ in a sequence
$\sigma$. For notational convenience we introduce the mutation
operator $\Delta_l: \mathbb{H}_2^L \to \mathbb{H}_2^L$ which flips the allele at locus $l$ and
is formally defined through \cite{Hwang2018}  
\begin{equation}
  \label{eq:Delta}
  (\Delta_l \sigma)_k = (1-\delta_{kl}) \sigma_k + \delta_{kl}(1-\sigma_k).
\end{equation}
The effect sizes of the two single mutations are given by $s_{i,j} = g(\Delta_{i,j} \sigma) - g(\sigma)$, and the pairwise epistatic
interaction $\epsilon_{ij}$ measures the deviation of the effect size
of the double mutant from the additive prediction $s_i + s_j$:
\begin{equation}
  \label{eq:epsilon2}
  \epsilon_{ij} = g(\Delta_i \Delta_j \sigma) - g(\sigma) - (s_i + s_j) = g(\Delta_i \Delta_j \sigma) + g(\sigma)
  - g(\Delta_i \sigma) - g(\Delta_j \sigma).
\end{equation}
We say that the mutations at positions $i$ and $j$ interact
epistatically in the \textit{background} $\sigma$ if $\epsilon_{ij}
\neq 0$, and the interaction is positive or negative according to the sign of
$\epsilon_{ij}$.  
Similar expressions can be written to quantify epistatic interactions of higher order \cite{Crona2017,Szendro2013,Poelwijk2016}.

The term \textit{universal epistasis} was introduced in \cite{Crona2023} to describe a global constraint on a fitness landscape
that relates the ordering between the fitness effects of a mutation in different backgrounds
to the inclusion relation (in the sense of section \ref{Sec:binary}) between the background gentoypes. Specifically, consider
two genotypes $\sigma$, $\sigma'$ with $\sigma' \subset \sigma$, as well as a set $\tau$ of sites not contained in $\sigma$.
Then the fitness landscape displays universal positive (negative) epistasis, if the fitness
effect of the composite mutational event $\sigma \to \sigma \cup \tau$
is always greater (smaller) than the effect of the mutation $\sigma' \to \sigma' \cup \tau$. In the present context
the condition of universal negative epistasis (UNE) is of primary interest, which can be formally written as
\begin{equation}
  \label{eq:UNE}
  g(\sigma \cup \tau) - g(\sigma) \leq g(\sigma' \cup \tau) - g(\sigma') \;\;
  \textrm{for} \; \textrm{any} \; \sigma' \subset \sigma, \; \tau \subset {\mathcal{L}} \setminus \sigma.
\end{equation}
In \ref{App} we show that the condition (\ref{eq:UNE}) holds for arbitrary $\sigma, \sigma',
\tau$ provided
it is satisfied for all epistatic interactions between pairs of single mutations in all
backgrounds, i.e., if
$\epsilon_{ij} \leq 0 \;\; \forall i,j,\sigma$ in (\ref{eq:epsilon2}).
In this special case, $d(\sigma,\sigma') = 1$ and the set $\tau$ in (\ref{eq:UNE}) consists of a single element.

The importance of the relation (\ref{eq:UNE}) lies in the observation (first noted in \cite{Das2020}) that a fitness
landscape with purely negative epistasis satisfies the accessibility property (AP) defined in section \ref{sec:AP}. Here   
we show how the AP follows from (\ref{eq:UNE}). Suppose that the genotype $\sigma$ is a fitness peak, and consider
a subset genotype $\sigma' \subset \sigma$. Then we need to show that the addition of any mutation $i \in \sigma \setminus \sigma'$
to $\sigma'$ increases fitness. Because $\sigma$ is a peak, we know that $g(\sigma) - g(\sigma \setminus \{i\}) > 0$
for any $i$, and using (\ref{eq:UNE}) it follows that
\begin{equation}
  \label{eq:proofAP}
  g(\sigma' \cup \{i\}) - g(\sigma') \geq g(\sigma) - g(\sigma \setminus \{i\}) > 0.
\end{equation}
The accessibility of the peak from superset genotypes can be proved in the same way.

\subsection{Fitness landscapes with an intermediate phenotype}
\label{Sec:intermediate}

A simple way to generate fitness landscapes with the UNE property (\ref{eq:UNE}) is to combine a linear genotype-phenotype
map with a suitably chosen nonlinear function that maps the phenotype
to fitness \cite{Manrubia2021}.
Fitness landscapes based on such a construction have been studied theoretically in the context
of Fisher's geometric model \cite{Fisher1958,Tenaillon2014,Hwang2017,Park2020}, but are also used in the analysis and interpretation of empirical
data sets \cite{Pokusaeva2019,Rokyta2011,Schenk2013,Otwinowski2018}.
Conceptually, they constitute an attempt to partially reinstate the intermediate phenotype
in the genotype-fitness map (figure \ref{fig:fig1}).

\begin{figure}[htb]
    \centering
    \includegraphics[width=0.7\textwidth]{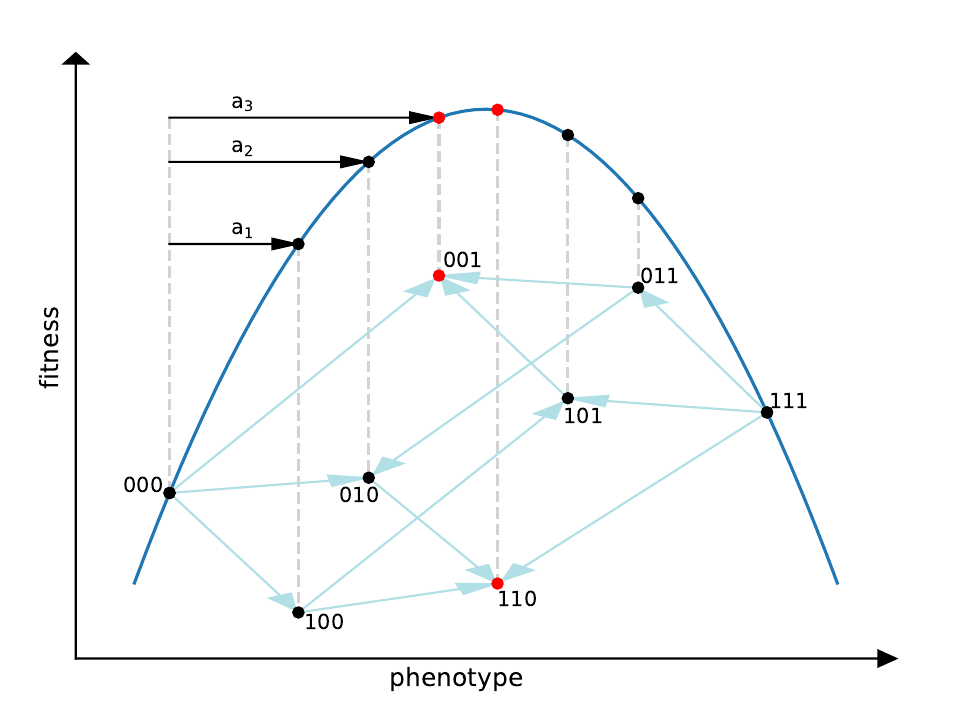}
    \caption{\textbf{Genotypic fitness landscape induced by a
        one-dimensional 
        nonlinear phenotype-fitness map.} 
      The figure illustrates the construction of a fitness landscape of the form (\ref{eq:UNEmodel}) with
      $L=3$. The fitness of a genotype depends nonlinearly on the
      corresponding phenotype value,
      which is a linear combination of the positive mutational effects $a_1,
      a_2, a_3$. The dashed horizontal lines connect the fitness
      values to the nodes of the 3-cube. Because the phenotype-fitness
    map is non-monotonic, the genotypic fitness landscape can have
    multiple peaks, which are marked in red.}
    \label{fig:FGM3}
\end{figure}

\begin{figure}[htb]
    \centering
    \includegraphics[width=0.7\textwidth]{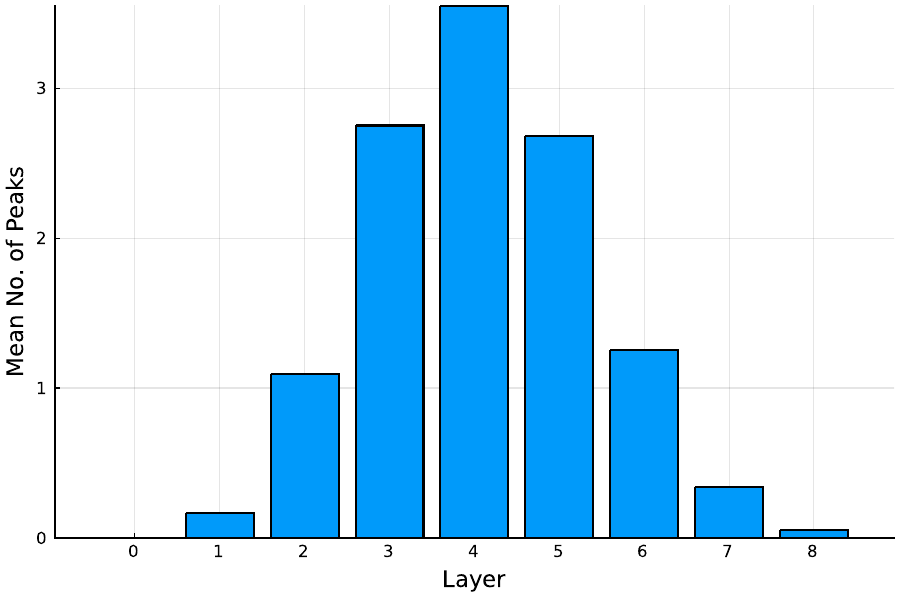}
    \caption{\textbf{Distribution of the number of fitness peaks in a landscape with universal negative epistasis.} 
      The figure shows the distribution of the number of fitness peaks across layers of constant mutation number $n_\sigma$ for
      a fitness landscape of the form (\ref{eq:UNEmodel}). The dimension is $L=8$, the coefficients $a_i > 0$ are chosen from an
      exponential distribution with unit mean, and the phenotype-fitness map is $\Phi(z) = - (z-L\langle a_i \rangle/2)^2 =
    -(z-4)^2$. The data constitute an average over 20000 realizations.}
    \label{fig:FGM1}
\end{figure}

\begin{figure}[htb]
    \centering
     \includegraphics[width=0.9\textwidth]{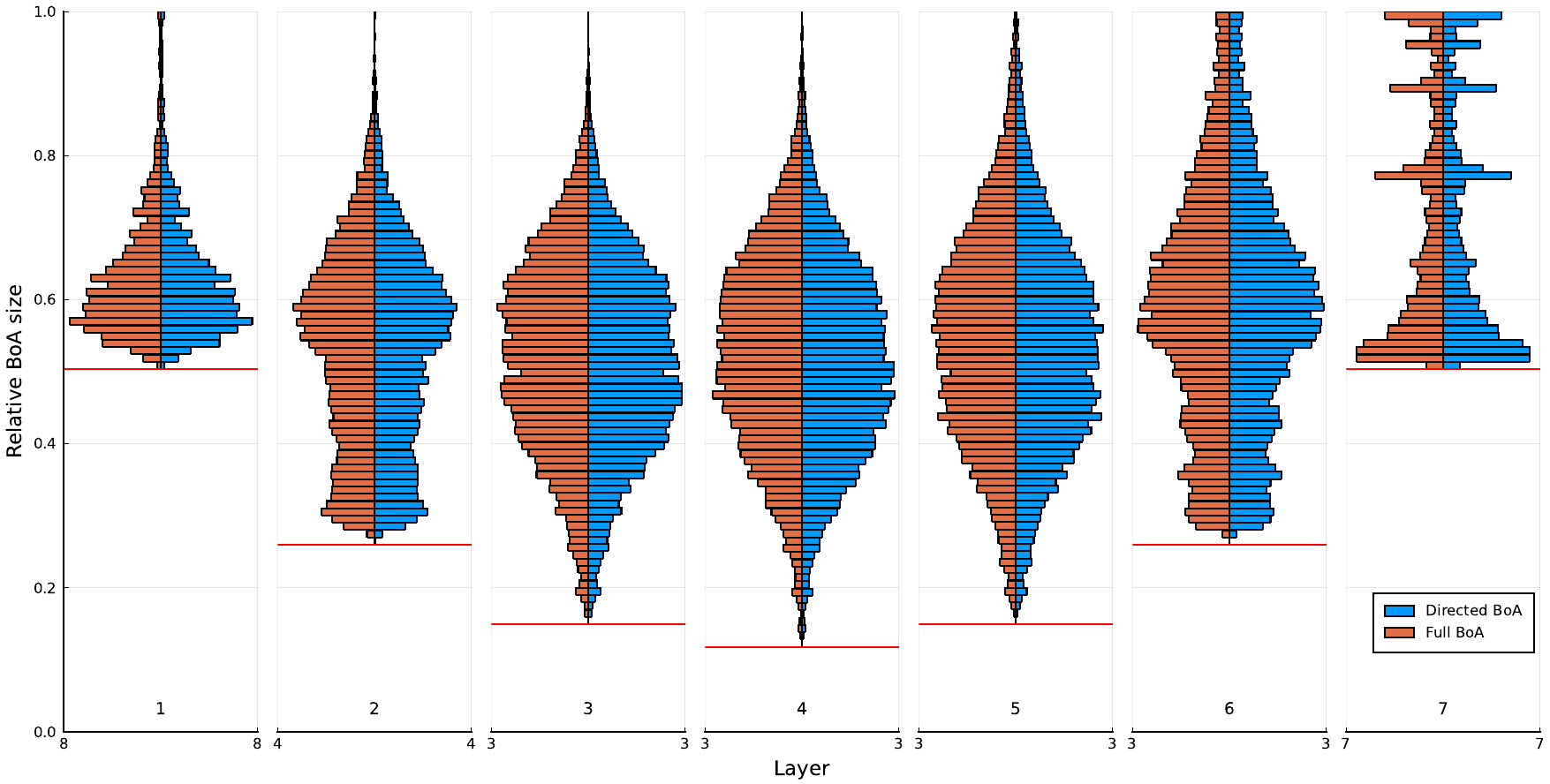}
    \caption{\textbf{Basins of attraction of fitness maxima in a
        landscape with universal negative epistasis.} The figure shows
      histograms for the relative size (\ref{eq:BoA_rel}) of the BoA's of
      fitness peaks for the model described in figure \ref{fig:FGM1}. The
      layout is the same as in figure \ref{fig:TIL_BoA}.  
    }
    \label{fig:FGM2}
\end{figure}

For the case of a one-dimensional intermediate phenotype $z$, the fitness function is of the form
\begin{equation}
  \label{eq:UNEmodel}
  g(\sigma) = \Phi[z(\sigma)] \;\; \textrm{with} \;\; z(\sigma) = \sum_{i=1}^L a_i \sigma_i. 
\end{equation}
To proceed, we make two further assumptions.
\begin{itemize}
\item[(i)] The function $\Phi$ is concave, that is,
  \begin{equation}
    \label{eq:concave}
    \Phi(x+a) - \Phi(x) < \Phi(y + a) - \Phi(y) \;\; \textrm{for} \; \textrm{any} \; a > 0, \; x > y.
  \end{equation}
\item[(ii)] The coefficients $a_i$ are positive.
\end{itemize}
For genotypes $\sigma, \sigma', \tau$ with $\sigma' \subset \sigma$ and $\tau \subset {\cal{L}}\setminus \sigma$ it follows
from (ii) that $z(\sigma) > z(\sigma')$ and $z(\sigma \cup \tau) = z(\sigma) + z(\tau)$,
$z(\sigma' \cup \tau) = z(\sigma') + z(\tau)$. Using additionally the concavity property (i), we find that  
\begin{eqnarray}
  \label{eq:concave2}
  g(\sigma \cup \tau) - g(\sigma) = \Phi[z(\sigma) + z(\tau)] - \Phi[z(\sigma)] < \nonumber \\
  \Phi[z(\sigma') + z(\tau)] - \Phi[z(\sigma')] = g(\sigma' \cup \tau) - g(\sigma')
\end{eqnarray}
and (\ref{eq:UNE}) is established. As an example, a landscape that saturates the bound (\ref{eq:APpeaks})
on the number of peaks is obtained
by taking $a_i \equiv \bar{a} > 0$ to be a constant and choosing $\Phi$ as a non-monotonic function with its maximum at
$x = \bar{a} \lfloor \frac{L}{2} \rfloor$.
Figure \ref{fig:FGM3} illustrates the construction for the general
case of unequal phenotypic effects $a_i$, and figures \ref{fig:FGM1} and \ref{fig:FGM2} show results of an exploratory numerical study of the
model using exponentially distributed coefficients $a_i$.
Comparing figures \ref{fig:TIL_BoA} and \ref{fig:FGM2}, we see that
the size distributions of the BoA's of fitness peaks have much less weight at values close to $S_\sigma^\textrm{rel} \approx 1$ in the present model. Importantly, the AP, which is satisfied for both models, does not fully specify the properties of the BoA's.

In fact, the positivity condition (ii) on the $a_i$ can be relaxed. For general (nonzero) coefficients we define a transformation
$\sigma \to \tilde{\sigma}$ through
 \begin{equation}
  \label{eq:sigmatilde}
 \tilde{\sigma}_i  = 
\begin{cases} \sigma_i & a_i > 0  \\
1-\sigma_i  & a_i < 0
\end{cases}
 \end{equation}
 and rewrite the linear phenotype as
 \begin{equation}
   \label{eq:transformz}
   z(\sigma) = \sum_{i=1}^L a_i \sigma_i = \sum_{i: a_i > 0} \vert a_i \vert \sigma_i - \sum_{i: a_i < 0} \vert a_i \vert \sigma_i  = 
   \sum_{i} \vert a_i \vert \tilde{\sigma_i} + \sum_{i: a_i < 0} a_i.
   \end{equation}
Thus $z$ is a linear function of the $\tilde{\sigma}$ with positive
coefficients, and the relation (\ref{eq:concave2}) holds in the transformed
coordinates.

These considerations can be applied to Fisher's geometric model with a
one-dimensional phenotype \cite{Hwang2017,Park2020}. In this model the coefficients
$a_i$ are random variables sampled from a symmetric distribution with
zero mean, and the phenotype-fitness map has a unique maximum at
$z=0$. This corresponds to a situation where
selection favors an intermediate value of a phenotype, a scenario that
makes good sense biologically \cite{Srivastava2022}. In \cite{Park2020} it was shown that the expected number of
fitness peaks in the one-dimensional model is of the order of
\begin{equation}
  \label{eq:NFisher}
  \mathbb{E}[N] \sim \frac{2^L}{L^{3/2}}
\end{equation}
for large $L$,
and the full distribution of $N$ was computed. Thus these fitness
landscapes are highly rugged. The arguments presented
above show that, provided the function $\Phi(z)$ is concave, the model
satisfies the subset-superset accessibility property. A simple example of a concave function is the
inverse parabola, $\Phi(z) \sim - z^2$. As explained in
\cite{Park2020}, with this choice Fisher's
geometric model is closely related to the antiferromagnetic Hopfield
model \cite{Nokura1998}, and it can also be mapped to the number partitioning problem
\cite{Ferreira1998}. This suggests that the AP holds for the optimization landscapes of these
problems as well. 

\section{Summary and outlook}
\label{Sec:Conclusions}

The dynamics of many complex systems can be described as a search process in a high-dimensional
landscape defined by some objective function such as energy or fitness. Probabilistic
models defined over discrete configuration spaces provide a versatile
framework for exploring the properties of these landscapes, and are of great current interest in diverse fields ranging from
evolutionary biology to deep learning \cite{Mezard2023}. In this article we have
reviewed a particular notion of landscape accessibility that originated
in the context of biological fitness landscapes. Detailed case studies
of two classical fitness landscape models, the HoC and the NK-model,
were summarized in sections \ref{Sec:random} and
\ref{Sec:structured}. Somewhat surprisingly, we find that accessible pathways
covering distances linear in the landscape dimension $L$ are likely to
exist even in the uncorrelated random HoC landscape, provided the
fitness gain covered by the path is large enough. At the same time, correlated landscapes of the NK-type are generally less, not
more accessible than random landscapes.

Section \ref{Sec:highly} describes a novel
class of fitness landscapes with a graded structure
which enables high ruggedness (many peaks) to coexist with high
accessibility. These landscapes are characterized by the
subset-superset accessibility property (AP), which couples the
accessibility of fitness peaks to the inclusion relations between
genotypes when interpreted as sets of loci.   
An immediate but striking consequence of the AP is a \textit{lower} bound on the size of the basin of attraction of
any fitness peak that grows exponentially with $L$, as $2^{L/2}$ [see equation
(\ref{eq:basin})]. The AP was originally found in a model of
tradeoff-induced fitness landscapes \cite{Das2020}, but it can also be
derived as a consequence of negative epistatic
interactions. Importantly, the minimal (necessary) conditions for the
AP to hold are not presently known.

We close by describing some problems for future research that need to
be addressed in order to apply the ideas presented here to large-scale
empirical data sets such as those in \cite{Papkou2023} and \cite{Westmann2023}. First, the
work on structured fitness landscapes should be generalized to sequence
spaces with more than two alleles \cite{Srivastava2023}. In
particular, the considerations for the AP-landscapes
discussed in section \ref{Sec:highly} need to be modified to account
for the geometry of multiallelic sequence spaces. Second, a better understanding of the correlations between
different quantifiers of landscape topography for single instances of
a probabilistic model is required. For example, how does the number and
height of fitness peaks correlate with the sizes of their basins of
attraction? While properties of BoA's remain unexplored even for the most basic fitness landscape models,
recent studies on the joint distribution of the number of
peaks and the frequency of different types of pairwise epistatic
interactions \cite{Oros2022,Saona2022,Riehl2022} have begun to shed light on related questions.

\paragraph*{Note added.} 
After submission of the revised version of the manuscript, we noticed that universal negative epistasis is equivalent to a property
of set functions known as submodularity \cite{Krause2014}. The consequences of this observation will be investigated elsewhere.

\ack
We are much indebted to Kristina Crona, Suman Das, Jasper Franke, Muhittin Mungan, Stefan Nowak and Benjamin Schmiegelt for their contributions to the results described here, and to Suman Das and Muhittin Mungan for comments on the manuscript.
Many thanks to Peter M\"orters for making us aware of the connection to submodularity. This work was
supported by DFG within the grants SFB 680, SFB 1310 and SPP 1590.

\appendix
\section{\label{App} Pairwise negative epistasis implies universal negative epistasis}

For the following considerations it is useful to rewrite the UNE condition (\ref{eq:UNE}) in the equivalent form \cite{Crona2023}
\begin{equation}
  \label{eq:UNE2}
  g(\upsilon \cup \upsilon') + g(\upsilon \cap \upsilon') \leq g(\upsilon) + g(\upsilon'),
\end{equation}
which is obtained from (\ref{eq:UNE}) by setting $\upsilon = \sigma$ and $\upsilon' = \sigma' \cup \tau$. In this formulation,
the condition holds for arbitrary pairs $\upsilon, \upsilon'$. We start from two genotypes at maximal distance,
$d(\upsilon, \upsilon') = L$, and show that (\ref{eq:UNE2}) is satisfied, provided it holds for pairs at strictly smaller
distance. By iterating the argument, it follows that it is sufficient to demand non-positive epistasis for all pairs
of genotypes at distance $d=2$. Geometrically, these pairs make up the 2-faces of the binary $L$-cube.

The maximal distance condition $d(\upsilon, \upsilon') = L$ implies that $\upsilon' = \bar{\upsilon}$ and
therefore $\upsilon \cup \upsilon' = {\cal{L}}$ and $\upsilon \cap \upsilon' = \emptyset$. We now choose a proper,
non-empty subset $\upsilon_1 \subset \upsilon'$, which can be done in $2^{\vert \upsilon' \vert}-2$ different ways.
If this is not possible because $\vert \upsilon' \vert < 2$, $\upsilon_1$ is chosen to be a subset of $\upsilon$ instead,
and the roles of $\upsilon'$ and $\upsilon$ are interchanged.  
Next define $\upsilon_2 \equiv \upsilon_1 \cup \upsilon$. Our strategy is to assume UNE for the pairs
\begin{equation}
  \label{eq:pairs}
  (\upsilon_1, \upsilon_1') \equiv (\upsilon_1, \upsilon) \;\;\; \textrm{and} \;\;\;
  (\upsilon_2, \upsilon_2') \equiv (\upsilon_2, \upsilon') = (\upsilon_1 \cup \upsilon, \upsilon').
\end{equation}
The following identities are easily checked:
\begin{equation}
  \upsilon_1 \cap \upsilon_1' = \emptyset, \;\; \upsilon_1 \cup \upsilon_1' = \upsilon_1 \cup \upsilon, \;\; \upsilon_2 \cap \upsilon_2' = \upsilon_1, \;\; \upsilon_2 \cup \upsilon_2' = {\cal{L}}.
\end{equation}
Moreover, making use of the general relation
\begin{equation}
  d(\chi, \chi') = d(\chi' \cup \chi, \chi' \cap \chi)
\end{equation}
for arbitrary pairs of genotypes $\chi, \chi'$, we find
\begin{equation}
  d(\upsilon_1, \upsilon_1') = d(\emptyset, \upsilon_1 \cup \upsilon) =
  \vert \upsilon_1 \cup \upsilon \vert < L, \;\; d(\upsilon_2, \upsilon_2') = d({\cal{L}},\upsilon_1) = L- \vert \upsilon_1 \vert < L
\end{equation}
because $\upsilon_1$ is a proper, non-empty subset of $\upsilon' = \bar{\upsilon}$. Thus the distances between the genotypes making
up the pairs (\ref{eq:pairs}) are strictly smaller than $L$. The UNE conditions for the pairs read
\begin{eqnarray}
  g(\upsilon_1 \cup \upsilon_1') + g(\upsilon_1 \cap \upsilon_1') \leq g(\upsilon_1) + g(\upsilon_1') \\
  g(\upsilon_2 \cup \upsilon_2') + g(\upsilon_2 \cap \upsilon_2') \leq g(\upsilon_2) + g(\upsilon_2').
\end{eqnarray}
Adding the two inequalities and cancelling identical terms on both
sides, the UNE condition (\ref{eq:UNE2}) for $(\upsilon, \upsilon')$
follows. Unless the pairs (\ref{eq:pairs}) already form 2-faces,
$d(\upsilon_1, \upsilon_1') = d(\upsilon_2, \upsilon_2') = 2$, the argument is repeated for the smaller subcubes spanned by
$(\upsilon_1 \cup \upsilon_1', \upsilon_1 \cap \upsilon_1')$ and
$(\upsilon_2 \cup \upsilon_2', \upsilon_2 \cap \upsilon_2')$.
The same proof applies to the case of universal positive epistasis considered in \cite{Crona2023}.

\section*{References}
\bibliographystyle{iopart-num.bst}
\bibliography{Krug_Oros}

\providecommand{\newblock}{}
\begin{thebibliography}{10}
\expandafter\ifx\csname url\endcsname\relax
  \def\url#1{{\tt #1}}\fi
\expandafter\ifx\csname urlprefix\endcsname\relax\def\urlprefix{URL }\fi
\providecommand{\eprint}[2][]{\url{#2}}

\bibitem{Wright1932}
Wright S 1932 The roles of mutation, inbreeding, crossbreeding, and selection
  in evolution {\em Proc. Sixth. Int. Cong. Genet.\/} vol~1 pp 356--366

\bibitem{Svensson2012}
Svensson E~I and Calsbeek R 2012 {\em The Adaptive Landscape in Evolutionary
  Biology\/} (Oxford University Press)

\bibitem{deVisser2014}
de~Visser J~A~G~M and Krug J 2014 {\em Nature Reviews Genetics\/} {\bf 15} 480

\bibitem{Hartl2014}
Hartl D~L 2014 {\em Current Opinion in Microbiology\/} {\bf 21} 51--57

\bibitem{Kondrashov2015}
Kondrashov D~A and Kondrashov F~A 2015 {\em Trends in Genetics\/} {\bf 31}
  24--33

\bibitem{Fragata2019}
Fragata I, Blanckaert A, Louro M~A~D, Liberles D~A and Bank C 2019 {\em Trends
  in ecology \& evolution\/} {\bf 34} 69--82

\bibitem{Manrubia2021}
Manrubia S, Cuesta J~A, Aguirre J, Ahnert S~E, Altenberg L, Cano A~V,
  Diaz-Uriarte P~C~R, Elena S~F, Garc\'{i}a-Mart\'{i}n J~A, Hogeweg P, Khatri
  B~S, Krug J, Louis A~A, Martin N~S, Payne J~L, Tarnowski M~J and Wei{\ss} M
  2021 {\em Physics of Life Reviews\/} {\bf 38} 55--106

\bibitem{Bank2022}
Bank C 2022 {\em Annual Review of Ecology, Evolution, and Systematics\/} {\bf
  53} 457--479

\bibitem{Provine1986}
Provine W~B 1986 {\em Sewall Wright and Evolutionary Biology\/} (The University
  of Chicago Press)

\bibitem{Agarwala2019}
Agarwala A and Fisher D~S 2019 {\em Theor. Pop. Biol.\/} {\bf 130} 13--49

\bibitem{Greenbury2022}
Greenbury S~F, Louis A~A and Ahnert S~E 2022 {\em Nature Ecology \&
  Evolution\/} {\bf 6} 1742--1752

\bibitem{Weinreich2005}
Weinreich D~M, Watson R~A and Chao L 2005 {\em Evolution\/} {\bf 59} 1165--1174

\bibitem{Poelwijk2007}
Poelwijk F~J, Kiviet D~J, Weinreich D~M and Tans S~J 2007 {\em Nature\/} {\bf
  445} 383

\bibitem{Carneiro2010}
Carneiro M and Hartl D~L 2010 {\em Proc. Natl. Acad. Sci. USA\/} {\bf 107}
  1747--1751

\bibitem{Franke2011}
Franke J, Kl\"ozer A, de~Visser J~A~G~M and Krug J 2011 {\em PLoS Computational
  Biology\/} {\bf 7} e1002134

\bibitem{Weinreich2006}
Weinreich D~M, Delaney N~F, DePristo M~A and Hartl D~L 2006 {\em Science\/}
  {\bf 312} 111--114

\bibitem{DePristo2007}
DePristo M~A, Hartl D~L and Weinreich D~M 2007 {\em Molecular biology and
  evolution\/} {\bf 24} 1608--1610

\bibitem{Lozovsky2009}
Lozovsky E~R, Chookajorn T, Brown K~M, Imwong M, Shaw P~J, Kamchonwongpaisan S,
  Neafsey D~E, Weinreich D~M and Hartl D~L 2009 {\em Proc. Natl. Acad. Sci.
  USA\/} {\bf 106} 12025--12030

\bibitem{Palmer2015}
Palmer A~C, Toprak E, Baym M, Kim S, Veres A, Bershtein S and Kishony R 2015
  {\em Nature Communications\/} {\bf 6} 7385

\bibitem{Bank2016}
Bank C, Matuszewski S, Hietpas R~T and Jensen J~D 2016 {\em Proc. Natl. Acad.
  Sci. USA\/} {\bf 113} 14085--14090

\bibitem{Wu2016}
Wu N~C, Dai L, Olson C~A, Lloyd-Smith J~O and Sun R 2016 {\em Elife\/} {\bf 5}
  e16965

\bibitem{Aguilar2017}
Aguilar-Rodr\'{i}guez J, Payne J~L and Wagner A 2017 {\em Nature Ecology \&
  Evolution\/} {\bf 1} 0045

\bibitem{Domingo2018}
Domingo J, Diss G and Lehner B 2018 {\em Nature\/} {\bf 558} 117--121

\bibitem{Pokusaeva2019}
Pokusaeva V~O, Usmanova D~R, Putintseva E~V, Espinar L, Sarkisyan K~S, Mishin
  A~S, Bogatyreva N~S, Ivankov D~N, Akopyan A~V, Avvakumov S~Y, Povolotskaya
  I~S, Filion G~J, Carey L~B and Kondrashov F~A 2019 {\em PLoS Genetics\/} {\bf
  15} e1008079

\bibitem{Moulana2022}
Moulana A, Dupic T, Phillips A~M, Chang J, Nieves S, Roffler A~A, Greaney A~J,
  Starr T~N, Bloom J~D and Desai M~M 2022 {\em Nature Communications\/} {\bf
  13} 7011

\bibitem{Papkou2023}
Papkou A, Garcia-Pastor L, Escudero J~A and Wagner A 2023 {\em Science\/} {\bf
  382} eadh3860

\bibitem{Westmann2023}
Westmann C~A, Goldbach L and Wagner A 2023 {\em bioRxiv preprint\/}

\bibitem{Schmiegelt2023}
Schmiegelt B and Krug J 2023 {\em Journal of Mathematical Biology\/} {\bf 86}
  46

\bibitem{Stadler1999}
Stadler P~F and Happel R 1999 {\em Journal of Mathematical Biology\/} {\bf 38}
  435--478

\bibitem{Altenberg2015}
Altenberg L 2015 {\em arXiv preprint arXiv:1508.07866\/}

\bibitem{deVisser2009}
de~Visser J~A~G~M, Park S~C and Krug J 2009 {\em American Naturalist\/} {\bf
  174} S15--S30

\bibitem{Crona2013}
Crona K, Greene D and Barlow M 2013 {\em Journal of Theoretical Biology\/} {\bf
  318} 1--10

\bibitem{Plotkin2011}
Plotkin J~B and Kudla G 2011 {\em Nature Reviews Genetics\/} {\bf 12} 32--42

\bibitem{Zwart2018}
Zwart M~P, Schenk M~F, Hwang S, Koopmanschap B, de~Lange N, van~de Pol L, Nga
  T~T~T, Szendro I~G, Krug J and de~Visser J~A~G~M 2018 {\em Heredity\/} {\bf
  121} 406--421

\bibitem{Kingman1978}
Kingman J~F 1978 {\em Journal of Applied Probability\/} {\bf 15} 1--12

\bibitem{Kauffman1987}
Kauffman S and Levin S 1987 {\em Journal of theoretical Biology\/} {\bf 128}
  11--45

\bibitem{Derrida1981}
Derrida B 1981 {\em Physical Review B\/} {\bf 24} 2613--2326

\bibitem{Crona2017}
Crona K, Gavryushkin A, Greene D and Beerenwinkel N 2017 {\em eLife\/} {\bf 6}
  e28629

\bibitem{Nowak2013}
Nowak S and Krug J 2013 {\em EPL\/} {\bf 101} 66004

\bibitem{Krug2021}
Krug J 2021 Accessibility percolation in random fitness landscapes {\em
  Probabilistic Structures in Evolution\/} (EMS Press) pp 1--22

\bibitem{Schmiegelt2023a}
Schmiegelt B 2023 {\em Structure and accessibility of fitness landscapes\/}
  Ph.D. thesis University of Cologne

\bibitem{Zagorski2016}
Zagorski M, Burda Z and Waclaw B 2016 {\em PLoS Computational Biology\/} {\bf
  12} e1005218

\bibitem{Hegarty2014}
Hegarty P and Martinsson A 2014 {\em Ann.~Appl.~Probab.\/} {\bf 24} 1375--1395

\bibitem{Berestycki2016}
Berestycki J, Brunet E and Shi Z 2016 {\em Bernoulli\/} {\bf 22} 653--680

\bibitem{Berestycki2017}
Berestycki J, Brunet E and Shi Z 2017 {\em ALEA, Lat. Am. J. Probab. Math.
  Stat.\/} {\bf 14} 45--62

\bibitem{Martinsson2018}
Martinsson A 2018 {\em Ann. Prob.\/} {\bf 46} 1004--1041

\bibitem{Martinsson2015}
Martinsson A 2015 {\em arXiv preprint arXiv:1501.02206\/}

\bibitem{Li2018}
Li L 2018 {\em J. Theor. Prob.\/} {\bf 31} 2072--2111

\bibitem{Kistler2020}
Kistler N and Schertzer A 2020 {\em arXiv preprint arXiv:2012.04076\/}

\bibitem{Szendro2013}
Szendro I~G, Schenk M~F, Franke J, Krug J and de~Visser J~A~G~M 2013 {\em
  Journal of Statistical Mechanics: Theory and Experiment\/} {\bf 2013} P01005

\bibitem{Das2020}
Das S~G, Direito S~O, Waclaw B, Allen R~J and Krug J 2020 {\em Elife\/} {\bf 9}
  e55155

\bibitem{Kauffman1989}
Kauffman S~A and Weinberger E~D 1989 {\em Journal of theoretical biology\/}
  {\bf 141} 211--245

\bibitem{Weinberger1991}
Weinberger E~D 1991 {\em Physical Review A\/} {\bf 44} 6399--6413

\bibitem{Hwang2018}
Hwang S, Schmiegelt B, Ferretti L and Krug J 2018 {\em Journal of Statistical
  Physics\/} {\bf 172} 226--278

\bibitem{Perelson1995}
Perelson A~S and Macken C~A 1995 {\em Proceedings of the National Academy of
  Sciences\/} {\bf 92} 9657--9661

\bibitem{Schmiegelt2014}
Schmiegelt B and Krug J 2014 {\em Journal of Statistical Physics\/} {\bf 154}
  334--355

\bibitem{Schmiegelt2016}
Schmiegelt B 2016 {\em Sign epistasis networks\/} Master's thesis University of
  Cologne

\bibitem{Aita2000}
Aita T, Uchiyama H, Inaoka T, Nakajima M, Kokubo T and Husimi Y 2000 {\em
  Biopolymers\/} {\bf 54} 64--79

\bibitem{Neidhart2014}
Neidhart J, Szendro I~G and Krug J 2014 {\em Genetics\/} {\bf 198} 699--721

\bibitem{Wolfinger2004}
Wolfinger M~T, Svrcek-Seiler W~A, Flamm C, Hofacker I~L and Stadler P~F 2004
  {\em Journal of Physics A\/} {\bf 37} 4731--4741

\bibitem{Franke2012}
Franke J and Krug J 2012 {\em Journal of Statistical Physics\/} {\bf 148}
  705--722

\bibitem{Das2022}
Das S, Krug J and Mungan M 2022 {\em Physical Review X\/} {\bf 12} 031040

\bibitem{Oros2022}
Oros D 2022 {\em Hypercubes, Peak Patterns and Universal Positive Epistasis\/}
  Master's thesis University of Cologne

\bibitem{Sperner1928}
Sperner E 1928 {\em Mathematische Zeitschrift\/} {\bf 27} 544--548

\bibitem{Haldane1931}
Haldane J 1931 {\em Mathematical Proceedings of the Cambridge Philosophical
  Society\/} {\bf 27} 137--142

\bibitem{Crona2023}
Crona K, Krug J and Srivastava M 2023 {\em Journal of Mathematical Biology\/}
  {\bf 86} 62

\bibitem{Regoes2004}
Regoes R~R, Wiuff C, Zappala R~M, Garner K~N, Baquero F and Levin B~R 2004 {\em
  Antimicrobial Agents and Chemotherapy\/} {\bf 48} 3670--3676

\bibitem{Rice2000}
Rice S~H 2000 The evolution of developmental interactions: Epistasis,
  canalization, and integration {\em Epistasis and the Evolutionary Process\/}
  (Oxford University Press) pp 82--98

\bibitem{Wolf2000}
Wolf J~B, {Brodie III} E~D and Wade M~J (eds) 2000 {\em Epistasis and the
  Evolutionary Process\/} (Oxford University Press)

\bibitem{Phillips2008}
Phillips P~C 2008 {\em Nature Reviews Genetics\/} {\bf 9} 855--867

\bibitem{Ferretti2016}
Ferretti L, Schmiegelt B, Weinreich D, Yamauchi A, Kobayashi Y, Tajima F and
  Achaz G 2016 {\em Journal of theoretical biology\/} {\bf 396} 132--143

\bibitem{Domingo2019}
Domingo J, Baeza-Centurion P and Lehner B 2019 {\em Annual Review of Genomics
  and Human Genetics\/} {\bf 20} 17.1--17.28

\bibitem{Krug2021a}
Krug J 2021 Epistasis and evolution {\em Oxford Bibliographies in Evolutionary
  Biology\/} (Oxford University Press)

\bibitem{Poelwijk2016}
Poelwijk F~J, Krishna V and Ranganathan R 2016 {\em PLoS Computational
  Biology\/} {\bf 12} e1004771

\bibitem{Fisher1958}
Fisher R~A 1958 {\em The genetical theory of natural selection\/} (Dover)

\bibitem{Tenaillon2014}
Tenaillon O 2014 {\em Annu. Rev. Ecol. Evol. Syst.\/} {\bf 45} 179--201

\bibitem{Hwang2017}
Hwang S, Park S~C and Krug J 2017 {\em Genetics\/} {\bf 206} 1049--1079

\bibitem{Park2020}
Park S~C, Hwang S and Krug J 2020 {\em Journal of Physics A: Mathematical and
  Theoretical\/} {\bf 53} 385601

\bibitem{Rokyta2011}
Rokyta D, Joyce P, Caudle S~B, Miller C, Beisel C~J and Wichman H~A 2011 {\em
  PLoS Genetics\/} {\bf 7} e1002075

\bibitem{Schenk2013}
Schenk M~F, Szendro I~G, Salverda M~L~M, Krug J and de~Visser J~A~G~M 2013 {\em
  Molecular biology and evolution\/} {\bf 30} 1779--1787

\bibitem{Otwinowski2018}
Otwinowski J, McCandlish D~M and Plotkin J~B 2018 {\em Proc. Natl. Acad. Sci.
  USA\/} {\bf 115} E7550--E7558

\bibitem{Srivastava2022}
Srivastava M and Payne J~L 2022 {\em PLoS Computational Biology\/} {\bf 18}
  e1010524

\bibitem{Nokura1998}
Nokura K 1998 {\em Journal of Physics A: Mathematical and Theoretical\/} {\bf
  31} 7447--7459

\bibitem{Ferreira1998}
Ferreira F and Fontanari J~F 1998 {\em Journal of Physics A: Mathematical and
  Theoretical\/} {\bf 31} 3417--3428

\bibitem{Mezard2023}
M{\'e}zard M 2023 {\em arXiv preprint arXiv:2309.06947\/}

\bibitem{Srivastava2023}
Srivastava M, Rozho\v{n}ov\'{a} H and Payne J~L 2023 {\em Journal of Physics A:
  Mathematical and Theoretical\/} {\bf 56} 455601

\bibitem{Saona2022}
Saona R, Kondrashov F~A and Khudiakova K~A 2022 {\em Bulletin of Mathematical
  Biology\/} {\bf 84} 74

\bibitem{Riehl2022}
Riehl M, Phillips R, Pudwell L and Chenette N 2022 {\em Journal of Physics A:
  Mathematical and Theoretical\/} {\bf 55} 434002

\bibitem{Krause2014}
Krause A and Golovin D 2014 Submodular function maximization {\em Tractability:
  Practical Approaches to Hard Problems\/} (Cambridge University Press,
  Cambridge, UK) pp 71--104

\end{thebibliography}
\end{document}